\documentclass[12pt]{article}
\usepackage[xdvi]{graphicx}
\usepackage{amssymb}
\newcommand{\bea}{\begin{eqnarray}}
\newcommand{\eea}{\end{eqnarray}}

\makeatletter
\@addtoreset{equation}{section}
\makeatother

\begin{document}
\begin{titlepage}
\begin{flushright}
OU-HET 615/2008
\end{flushright}

\vspace{25ex}

\begin{center}
{\Large\bf 
Operators in ultraviolet completions 
for collective symmetry breaking
}
\end{center}

\vspace{1ex}

\begin{center}
{\large 
Nobuhiro Uekusa}
\end{center}
\begin{center}
{\it Department of Physics, 
Osaka University \\
Toyonaka, Osaka 560-0043
Japan} \\
\textit{E-mail}: uekusa@het.phys.sci.osaka-u.ac.jp
\end{center}


\vspace{3ex}

\begin{abstract}

We investigate the Lagrangian terms of
scalar-gauge interactions in a weakly-coupled gauge theory
beyond the ultraviolet momentum cutoff
of models where the collective symmetry breaking
mechanism protects the Higgs mass squared against
having quadratic divergence.
We also propose a five-dimensional
model in which the gauge symmetry breaking 
is achieved by boundary conditions consistent
with local gauge transformations. 
It is shown that the scalar-gauge operators 
in the two models differ in the absence of quadratic divergence.

\end{abstract}
\end{titlepage}

\section{Introduction}

Quadratic divergence is one of the fundamental problems
of the Standard Model of particle physics.
The structure of quantum loop corrections is different
between scalars and fermions.
The self-energy of fermion is proportional to
the fermion mass. By a dimensional counting, it only
has logarithmic divergence at one loop.
It is chiral symmetry that protects fermions from 
having quadratic divergence.
Scalar fields do not have such a symmetry in the Standard Model. 
Therefore applying symmetry principle to scalar fields has been
a clue to understand physics beyond the Standard Model. 
Supersymmetry connects fermions and bosons so that
chiral symmetry is applied~%
\cite{Dimopoulos:1981zb,Sakai:1981gr}.
If scalars were components of gauge fields
which have gauge symmetry to prevent quadratic divergence,
it is formulated 
in a gauge-Higgs unification scenario~%
\cite{Hosotani:1983xw,Hosotani:1988bm,Hatanaka:1998yp}.
A feature of supersymmetry and gauge-Higgs scenarios is that
there are many partners: superparticles and
Kaluza-Klein modes.
In obtaining no quadratic divergence with symmetry principle,
it is not necessary to rely on symmetry for
fields with different spins than scalars.
If scalar fields were Nambu-Goldstone bosons in broken global 
symmetry, they are exactly massless and
it is led to little Higgs scenarios~%
\cite{ArkaniHamed:2001nc,ArkaniHamed:2002qy}.

In little Higgs scenarios, if gauge and Yukawa
couplings are vanishing,
Higgs fields have derivative couplings
and they have shift symmetry so that
there is no potential.
The scalar fields are described in a non-linear sigma
model.
The assumption that the global symmetry is
explicitly broken only when two or more couplings
are non-vanishing leads to a potential
with at most logarithmic divergence.
This collective symmetry breaking mechanism has been 
studied in many papers~%
\cite{Kaplan:2003uc,Cheng:2003ju,
Low:2004xc,Schmaltz:2005ky,
Asano:2006nr,Matsumoto:2008fq}.
In the littlest Higgs model, the resulting operator of
scalar-gauge interactions is, for example,
\bea
   W_{1\mu} W_2^{\,\mu} h^\dag h ,
   \label{basic}
\eea
where $h$ is the Higgs field and
$W_{1\mu}$ and $W_{2\mu}$ are $[\textrm{SU}(2)]^2$ gauge bosons.
At one loop, the gauge bosons 
do not produce the quadratic divergence for
the Higgs mass squared through the interaction (\ref{basic}),
as will be explained with diagrams in Section~\ref{rr}.
While the collective symmetry breaking mechanism requires two or
more couplings,
such a group as $[\textrm{SU}(2)]^2$ could be
a subgroup of a single group.
It should be clarified whether
operators such as Eq.~(\ref{basic}) can be derived
in a renormalizable gauge theory with a single large group broken to 
two or more subgroups.

Recently, in Ref.~\cite{Csaki:2008se}
a weakly-coupled renormalizable ultraviolet completion of a little 
Higgs scenario was proposed.
It was claimed that heavy modes are integrated out
and that the remaining theory is a non-linear sigma model due to the 
Coleman-Wess-Zumino theorem~\cite{Coleman:1969sm,Callan:1969sn}.
In the model,
the Higgs mass squared receives radiative correction 
of the order of 100~GeV and the non-linear sigma model
has the decay constant $f\sim 1$~TeV and 
the ultraviolet momentum cutoff $\Lambda\sim 10$~TeV.
Also, in an explicit example, several 
effective couplings with the form (\ref{basic}) were shown.
To pursue this possibility of high energy theory for 
no quadratic divergence as an extension of the collective 
symmetry breaking mechanism,
it would be important to examine more couplings.

We give all the couplings of
scalar-gauge interactions in a weakly-coupled gauge theory
beyond the ultraviolet momentum cutoff
of models where the collective symmetry breaking
mechanism protects the Higgs mass squared against
having quadratic divergence.
As another aspect, we study a possibility of gauge symmetry breaking 
by boundary conditions 
as the first step to utilize dynamical symmetry breaking in a gauge-Higgs unification.
We propose a five-dimensional model in which symmetry breaking
is achieved by boundary conditions consistent
with local gauge transformations.
To explain consistent boundary conditions,
we give some examples of inconsistent boundary conditions.
It is found that there is a difference of the structure between operators in the four-dimensional and five-dimensional models.

In the next section, the setup of a renormalizable 
gauge theory at higher energies than an ultraviolet momentum
cutoff of models with
the collective symmetry breaking mechanism
and the resulting couplings are given.
In Section \ref{bclrhs}, gauge symmetry breaking
by boundary conditions is examined. 
The couplings relevant to that of the model in
Section~\ref{rr} are given.
The character of the operators based on 
properties of symmetry breaking
is discussed in Section~\ref{interpret}.
Conclusion is made in Section~\ref{conc}.

\section{Emergence of operators with no quadratic divergence
 \label{rr}}

\begin{figure}[h]
\begin{center}
\includegraphics[width=3cm]{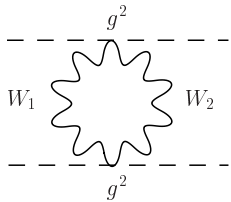}
\caption{The Higgs mass correction through 
the scalar-gauge interaction (\ref{basic2}). \label{fig1}}
\end{center}
\end{figure}

\begin{figure}[h]
\begin{center}
\setlength{\unitlength}{5.0mm}
\begin{picture}(10,5)(-2,0)
\put(-5,0){\includegraphics[width=3cm]{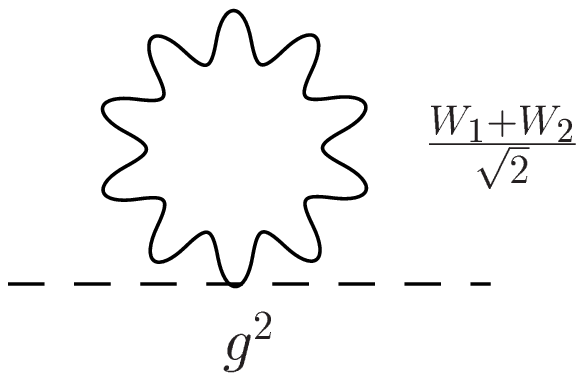}} 
\put(3,2){$+$}
\put(5,0){\includegraphics[width=3cm]{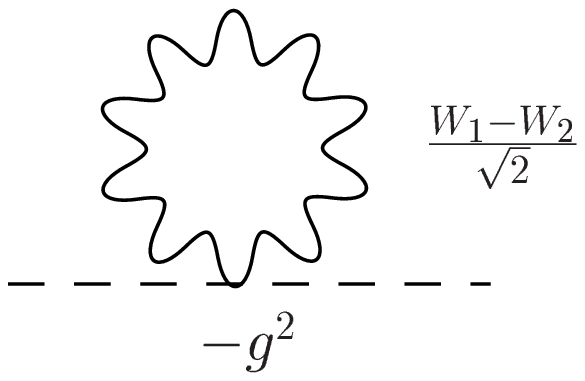}}
\end{picture}
\caption{The cancellation of quadratic divergence. \label{fig2}}
\end{center}
\end{figure}

We study a renormalizable gauge theory with a single group
which is broken by the vacuum expectation values
of scalar fields, aiming for having
scalar-gauge interactions of the form
\bea
   g^2 W_{1\mu} W_2^{\,\mu} h^\dag h ,
   \label{basic2}
\eea
where a gauge coupling is denoted as $g$.
For the interaction (\ref{basic2}), 
the one-loop diagram for
the Higgs mass squared is shown in Fig.~\ref{fig1}.
The divergence is at most logarithmic.
Since Eq.~(\ref{basic2}) is written as
\bea 
    g^2 \left({W_{1\mu}+W_{2\mu}\over \sqrt{2}}\right)^2 h^\dag h
  -g^2 \left({W_{1\mu}-W_{2\mu}\over \sqrt{2}}\right)^2 h^\dag h ,
   \label{basic3}
\eea
the absence of the quadratic divergence for the Higgs mass
squared is also regarded as a cancellation
as shown in Fig.~\ref{fig2}.
In little Higgs models, the collective symmetry breaking
mechanism can produce the operator (\ref{basic2}).
In such a model, the self-interaction of the Higgs fields obeys
a non-linear sigma model.
The Lagrangian terms of expansion at lower order are
\bea
   |\partial_\mu h|^2
      +{1\over f^2}|\partial_\mu h|^2 h^\dag h .
\eea  
The quadratic divergence of one-loop contribution to
the kinetic term is 
$\Lambda^2/(16\pi^2f^2)$,
where $\Lambda$ is the ultraviolet momentum cutoff.
For the non-linear sigma model to be viable,
the cutoff has the upper bound
$\Lambda < 4\pi f$.
From Figures \ref{fig1} and \ref{fig2},
the dominant correction to the Higgs mass squared
is 
\bea
   {g^4\over 16\pi^2} f^2 \log \left({\Lambda^2\over f^2}\right)
  .
\eea
If the correction is of the order of 100~GeV,
the decay constant is obtained as
$f\sim 1$~TeV and then $\Lambda\sim 10$~TeV.
Thus our starting theory is defined at higher
scales than 10~TeV.

We choose the single gauge group as SU(5).
The gauge symmetry is broken by the vacuum expectation
values of two types of scalar fields.
We assume that these two sectors 
yield separate gauge symmetry breakings
\bea
\textrm{SU}(5)\to [\textrm{SU(2)}\times\textrm{U(1)}]^2 ,
\qquad
\textrm{SU(5)$\to$SO(5)} .
\eea 
The Higgs fields are included in the sector of the
breaking SU(5)$\to$SO(5).
The scalar fields in the other sector are decoupled
to the Higgs fields.
The scalar fields we discuss are only in the 
sector of SU(5)$\to$SO(5).

\subsubsection*{Higgs mechanism}

Here we write down the Higgs mechanism of the symmetry breaking
SU(5)$\to$SO(5).
The scalar field responsible for this breaking is 
a scalar \underline{15} which transforms
as a complex symmetric matrix 
$S \to S'=U S U^T$
under SU(5).
The dynamics of $S$ is governed by a Lagrangian
invariant under global SU(5)$\times$U(1) and spacetime 
transformations.
The SU(5)$\times$U(1)-invariant potential is written as
\bea
   V = -M^2 \textrm{Tr} \left[SS^\dag\right]
     +\lambda_1 (\textrm{Tr}\left[SS^\dag\right])^2
     +\lambda_2 \textrm{Tr}[(SS^\dag)^2] ,
  \label{pot}
\eea
where $M$, $\lambda_1$ and $\lambda_2$ are the coupling constants.
The stationary condition leads to
\bea
  S^\dag \left[(-M^2 +2\lambda_1 \textrm{Tr}SS^\dag) \textrm{\bf 1}_5
     +2\lambda_2 SS^\dag \right]=0 .
\eea
The nonzero expectation value is given by
$SS^\dag =  f_S^2 \textrm{\bf 1}_5$ (for $10\lambda_1 + 2\lambda_2 \neq 0$),
which is proportional to the identity 
matrix. This means symmetry breaking to O(5).
Here the decay constant is defined as
$f_S^2 \equiv M^2/(10\lambda_1 +2\lambda_2)$.
For the vacuum expectation value
\bea
   \langle S\rangle =
     f_S \left(\begin{array}{ccc}
       &&  {\bf 1}_2 \\
       &1 &  \\
       {\bf 1}_2 && \\
     \end{array}\right) ,
      \label{vev}
\eea
the global SU(5)$\times$U(1) is broken to SO(5).

The vacuum expectation value (\ref{vev}) generates 
the masses of gauge bosons.
The gauge interaction of the scalar $S$ is included in
\bea
   \textrm{${1\over 8}$}
      \textrm{Tr} (D_\mu S)(D^\mu S)^\dag ,
\quad \textrm{with} \quad
    D_\mu S =\partial_\mu S
    -ig A_\mu^a T^a S
    - ig A_\mu^a S T^{a}{}^T ,
    \label{covs}
\eea
where properties of SU(5) generators are summarized in
App.~\ref{ap:su5g}.
Then the mass terms are obtained as
\bea
 &&
  \textrm{${1\over 8}$}g^2 f_S^2 \bigg[
   (A^1 +A^{22})^2
  + (A^2 -A^{23})^2
  + (A^4 +A^{13})^2
  + (A^5 +A^{14})^2
\nonumber
\\
 &&
  + (A^6 +A^{20})^2
  + (A^7 +A^{21})^2
  + 2A^9 A^9 
  + 2A^{10} A^{10}
\nonumber
\\
 &&
  + (A^{11} +A^{16})^2
  + (A^{12} -A^{17})^2
  + 2A^{18} A^{18}
  + 2A^{19} A^{19}
\nonumber
\\
 &&
  + A^3 A^3 + \textrm{${5\over 3}$} A^8 A^8
  + \textrm{${7\over 12}$} A^{15} A^{15}
  +\textrm{${3\over 4}$} A^{24} A^{24}
  -\textrm{${\sqrt{6}\over 2}$} A^3 A^{15}
\nonumber
\\
 &&
  + \textrm{${\sqrt{10}\over 2}$} A^3 A^{24}
  - \textrm{${5\sqrt{2}\over 6}$} A^8 A^{15}
  - \textrm{${\sqrt{30}\over 6}$} A^8 A^{24}
  - \textrm{${\sqrt{15}\over 6}$} A^{15} A^{24}
  \bigg] .
\eea
For the neutral sector $a=3,8,15,24$, the mass matrix is
\bea
 \textrm{${1\over 8}$} g^2 f_S^2
   \left(\begin{array}{cccc}
       1 & 0 & -{\sqrt{6}\over 4} & {\sqrt{10}\over 4} \\
       0 & {5\over 3} & -{5\sqrt{2}\over 12} & -{\sqrt{30}\over 12} \\
       -{\sqrt{6}\over 4} & -{5\sqrt{2}\over 12} & {7\over 12}
         & -{\sqrt{15}\over 12} \\
       {\sqrt{10}\over 4} & -{\sqrt{30}\over 12} &
         -{\sqrt{15}\over 12} & {3\over 4} \\
     \end{array} \right) .
\eea
The eigenvalues $m_A^2$ and eigenvectors are given by
\bea
  \left(\begin{array}{c}
     1 \\
     0 \\
     {\sqrt{6}\over 4} \\
     -{\sqrt{10}\over 4} \\
     \end{array}\right) ,
 \left(\begin{array}{c}
     0 \\
     {\sqrt{3}\over 3} \\
     {5\sqrt{6}\over 12} \\
     {\sqrt{10}\over 4} \\
     \end{array}\right) , ~ \textrm{for}~ m^2_A =0 ,
~
     \left(\begin{array}{c}
       1 \\
       0 \\
       -{\sqrt{6}\over 4} \\
       {\sqrt{10}\over 4} \\
     \end{array}\right) ,
   \left(\begin{array}{c}
       0 \\
     {5\sqrt{3}\over 3} \\
     -{5\sqrt{6}\over 12} \\
     -{\sqrt{10}\over 4} \\
     \end{array}\right) , ~
    \textrm{for} ~m^2_A=
  \textrm{${1\over 4}$}g^2 f_S^2 .
\eea 
The generators for massless fields are
\bea
  && T_{\overline{1}}=\textrm{${1\over \sqrt{2}}$}
  (T_1 -T_{22})  ,
\quad 
   T_{\overline{2}}=\textrm{${1\over \sqrt{2}}$}
   (T_2 +T_{23}), 
\quad
    T_{\overline{3}}=\textrm{${1\over \sqrt{2}}$}
  (T_4 -T_{13}) ,
\quad
   T_{\overline{4}}=\textrm{${1\over \sqrt{2}}$}
  (T_5 -T_{14}) ,   
\nonumber
\\
 &&   T_{\overline{5}}=\textrm{${1\over \sqrt{2}}$}
  (T_6 -T_{20}),
 \quad
       T_{\overline{6}}=\textrm{${1\over \sqrt{2}}$}
  (T_7 -T_{21}) ,
  \quad
    T_{\overline{7}}=\textrm{${1\over \sqrt{2}}$}
  (T_{11} -T_{16})  ,
  \quad
        T_{\overline{8}}=\textrm{${1\over \sqrt{2}}$}
  (T_{12} -T_{17}) ,
\nonumber
\\
  &&   T_{\overline{9}}
    = \textrm{${1\over \sqrt{2}}$}(T_3 +\textrm{${\sqrt{6}\over 4}$}T_{15}
      -\textrm{${\sqrt{10}\over 4}$}T_{24})  ,
  \quad
  T_{\overline{10}}
    = \textrm{${1\over \sqrt{2}}$}(\textrm{${\sqrt{3}\over 3}$}
   T_8 +\textrm{${5\sqrt{6}\over 12}$}T_{15}
      +\textrm{${\sqrt{10}\over 4}$}T_{24})  .
  \label{so5l}
\eea
These generators are given in a matrix form in Eq.~(\ref{matrixform}).
The generators
\bea
    T_{\overline{a}}  \left(\begin{array}{ccc}
        && {\bf 1}_2 \\
        &1& \\
        {\bf 1}_2 && \\
        \end{array}\right) ,
\eea
are a set of 5$\times$5 antisymmetric matrices.
They form SO(5) algebra.

\subsubsection*{Scalar fluctuations}

Next we identify scalar fields around the vacuum expectation value.
Fluctuations around the vacuum expectation value (\ref{vev}) are parameterized with 15 complex fields as
\bea
   S=\langle S\rangle + \bar{S} , 
   \qquad \textrm{with} \qquad
   \bar{S} = \left(\begin{array}{ccccc}
      A & B & C & D & E \\
      B^T & F & G & H & J \\
      C^T & G^T & K & L & M \\
      D^T & H^T & L^T & N & P \\
      E^T & J^T & M^T & P^T &Q \\
   \end{array}\right) .
  \label{fluct}
\eea
The fluctuation can be written as  
\bea
 \bar{S} = iN + R , \quad \textrm{with} \quad
   N=\left(\begin{array}{ccc}
      \phi & h & \chi \\ 
      h^T & K_i & h^\dag \\ 
      \chi^T & h^* & \phi^\dag \\
     \end{array}\right) , 
 \quad
 R =\left(\begin{array}{ccc}
    \Phi & H & X \\ 
    H^T & K_r & H^\dag \\ 
    X^T & H^* & \Phi^\dag \\ 
    \end{array} \right) .
\eea
which are composed of the two 2$\times$2 complex symmetric tensors
$\phi$ and $\Phi$, the two complex doublet fields $h$ and $H$,
the two 2$\times$2 Hermite tensors $\chi$ and $X$ and
the two real fields $K_i$ and $K_r$.
In terms of the fields in Eq.~(\ref{fluct}),
each matrix is given as
\bea
    \phi &\!\!\!=\!\!\!&
    \left(\begin{array}{cc}
      2\phi_a  & \sqrt{2} \phi_b \\
      \sqrt{2}\phi_b^T & 2\phi_c \\
      \end{array}\right)
 \equiv
 {i\over 2} \left( \begin{array}{cc}
       -A +N^\dag & -B + P^* \\
       -B^T + P^\dag & -F + Q^\dag \\
      \end{array}\right) ,
\\
    \Phi &\!\!\!=\!\!\!& 
  \left(\begin{array}{cc}
    2\bar{A} & \sqrt{2}\bar{B} \\
    \sqrt{2}\bar{B}^T & 2\bar{F} \\
    \end{array} \right)
   \equiv
{1\over 2} \left( \begin{array}{cc}
       A +N^\dag & B + P^* \\
       B^T + P^\dag & F + Q^\dag \\
      \end{array}\right) , 
\\
  h &\!\!\!=\!\!\!& 
   \sqrt{2} \bar{h} 
  \equiv
   {i\over 2} \left(\begin{array}{c}
     -C +L^\dag \\
     -G + M^\dag \\
     \end{array} \right) , \quad
  H = \sqrt{2} \bar{H} \equiv
   {1\over 2} \left(\begin{array}{c}
     C +L^\dag \\
     G + M^\dag \\
     \end{array} \right) ,
\\     
  K &\!\!\!=\!\!\!& 2i\bar{K}_i + 2\bar{K}_r \equiv iK_i + K_r ,
\\
  \chi &\!\!\!=\!\!\!&
  \sqrt{2} \left(\begin{array}{cc}
    \bar{D}_i & \chi_B \\
    \chi_B^\dag & \bar{J}_i \\
    \end{array}\right)
  \equiv
 {i\over 2} \left(\begin{array}{cc}
     -D + D^\dag & -E + H^\dag \\
     E^\dag - H & -J + J^\dag \\
      \end{array} \right) ,
\\
   X &\!\!\!=\!\!\!&
  \sqrt{2} \left(\begin{array}{cc}
    \bar{D}_r & \bar{E} \\
    \bar{E}^\dag & \bar{J}_r \\
    \end{array}\right)
     \equiv
 {1\over 2} \left(\begin{array}{cc}
     D + D^\dag & E + H^\dag \\
     E^\dag + H & J + J^\dag \\
      \end{array} \right) .
  \label{abcd}
\eea
The notation of the fields
with a bar over a letter such as $\bar{h}$ 
stand for the canonical normalization of the kinetic terms
\bea
 &&  \textrm{${1\over 8}$} \textrm{Tr} (\partial_\mu S)(\partial^\mu S^\dag)
\nonumber
\\
  &\!\!\!=\!\!\!&
   \textrm{${1\over 4}$} \textrm{Tr}
    \left[ (\partial_\mu \phi)(\partial^\mu \phi^\dag)
     +(\partial_\mu \chi)(\partial^\mu \chi^\dag)
   +(\partial_\mu \Phi)(\partial^\mu \Phi^\dag)
     +(\partial_\mu X)(\partial^\mu X^\dag)\right]
\nonumber
\\
  && +(\partial_\mu \bar{h}^\dag)(\partial^\mu \bar{h})
     +\textrm{${1\over 2}$}(\partial_\mu \bar{K}_i)
   ( \partial^\mu \bar{K}_i) 
   +(\partial_\mu \bar{H}^\dag)(\partial^\mu \bar{H})
     +\textrm{${1\over 2}$}(\partial_\mu \bar{K}_r)
   ( \partial^\mu \bar{K}_r)  .
 \label{kins}
\eea 
Substituting Eq.~(\ref{fluct}) into the potential (\ref{pot}) leads to
\bea
  V=
\lambda_1 \left[
  \textrm{Tr} (\langle S\rangle \bar{S}^\dag
  +\bar{S}\langle S\rangle^\dag )
   +\textrm{Tr} (\bar{S}\bar{S}^\dag) \right]^2
  +\lambda_2 \textrm{Tr}
  \left[ (\langle S\rangle \bar{S}^\dag
  +\bar{S}\langle S\rangle^\dag
  +\bar{S}\bar{S}^\dag)^2 \right] .
  \label{simppot}
\eea  
up to constant terms.
Because $\bar{S}\bar{S}^\dag$ is itself quadratic terms of fields,
the quadratic terms of scalar fields arise from
$\lambda_1 [\textrm{Tr}(\langle S\rangle \bar{S}^\dag
+\bar{S}\langle S\rangle^\dag)]^2$ 
and $\lambda_2 \textrm{Tr}[(\langle S\rangle \bar{S}^\dag
+\bar{S}\langle S\rangle^\dag)^2 ]$.
They include the operator
\bea
  \bar{S}\langle S\rangle^\dag
  +\langle S\rangle \bar{S}^\dag
   =2f_S
   \left(\begin{array}{ccc}
       X & H & \Phi  \\
       H^\dag & K_r & H^T \\
       \Phi^\dag & H^* & X^T \\
       \end{array}
       \right) 
   =2R\langle S \rangle 
 .
     \label{point}
\eea
From this equation,
it is seen that the fields in the matrix $R$
have mass terms.
Thus 15 real components are massive.
The number of massless components is $30-15=15$
which is equivalent to the number of broken generator SU(5)$\times$U(1)/SO(5).
The potential (\ref{simppot}) also includes the operator
\bea
   \bar{S}\bar{S}^\dag =(iN+R)(-iN^\dag +R^\dag)
   =NN^\dag + i(NR^\dag -RN^\dag) + RR^\dag .
    \label{ssbar}
\eea
With a symbolic use of $N,R$,
the structure of the potential is
\bea
  V &\!\!\!\sim \!\!\!&
    (R + N^2 + NR + R^2)^2 .
\eea
By power of $R$, it is classified as
\bea
  V \sim N^4, (N^2 R, N^3 R), (R^2 ,N R^2 , N^2 R^2),
   (R^3, NR^3) ,R^4.
   \label{class}
\eea
For effective vertices obtained by integrating out heavy fields,
lower $R$ terms 
$V \sim N^4$, $N^2 R$, $N^3 R$, $R^2$ are dominant. 
We obtain each term as follows:
The $N^4$ term is
\bea
   V_{N^4} &\!\!\!=\!\!\!& \lambda_1 \left[
     \textrm{Tr} NN^\dag \right]^2
     +\lambda_2 \textrm{Tr} \left[(NN^\dag)^2\right] 
\nonumber
\\
  &\!\!\!=\!\!\!&
     \lambda_1 \left( 2\textrm{Tr} (\phi\phi^\dag) 
        +2\textrm{Tr} (\chi^2) + K_i^2 + 4 h^\dag h \right)^2
\nonumber
\\
  &&
    + \lambda_2
      \left( 2\textrm{Tr} \left[(\phi\phi^\dag + hh^\dag
        +\chi^2 )^2 \right]
      +4 (\phi h^* + hK_i + \chi h)^\dag
         (\phi h^* + hK_i + \chi h)
       \right.
\nonumber
\\
 &&  \left.
   +2\textrm{Tr} \left[
    (\phi\chi^T + hh^T + \chi \phi)( \phi \chi^T + hh^T
  +\chi \phi)^\dag \right]
   +(2h^\dag h + K_i^2 )^2 \right)
 .      
\eea 
The $N^2 R$ term is
\bea
   V_{N^2 R} &\!\!\!=\!\!\!& 2\lambda_1 \textrm{Tr}
     (2R\langle S\rangle^\dag) \textrm{Tr}(NN^\dag)
   +2\lambda_2 \textrm{Tr} (2R \langle S\rangle^\dag NN^\dag) 
\nonumber
\\
   &\!\!\!=\!\!\!&
     4f_S \lambda_1 (2\textrm{Tr}(X) + K_r)
       (2\textrm{Tr}(\phi \phi^\dag) +2 \textrm{Tr} (\chi^2)
         +K_i^2 + 4 h^\dag h )
\nonumber
\\
  &&+ 4f_S \lambda_2 \left(
    2\textrm{Tr} \left[X(\phi \phi^\dag + hh^\dag + \chi^2)\right]
  \right.
\nonumber
\\
 &&
   +\textrm{Tr} \left[\Phi (2\chi^T \phi^\dag +h^*h^\dag)
           + (\Phi(2\chi^T \phi^\dag + h^* h^\dag))^* \right]
\nonumber
\\
  &&
    + 2H^T (\chi^T h^* +h^* K_i + \phi^\dag h)
    +\left(2H^T (\chi^T h^* +h^* K_i + \phi^\dag h)\right)^*
\nonumber
\\
  && \left.
  +K_r (2h^\dag h+K_i^2 )\right) .
\label{vn2r}
\eea
The $N^3 R$ term is $V_{N^3 R} = 0$.
The $R^2$ term is
\bea
   V_{R^2} &\!\!\!=\!\!\!& \lambda_1 \left[\textrm{Tr} (2R\langle S\rangle)\right]^2
    +\lambda_2 \textrm{Tr} \left[(2R\langle S\rangle)^2\right]  
\nonumber
\\
 &\!\!\!=\!\!\!&
   4f_S^2 \lambda_1 (2\textrm{Tr}(X)+K_r)^2 
   + 4f_S^2 \lambda_2 \left(
      2\textrm{Tr}(X^2) +2\textrm{Tr} (\Phi\Phi^\dag) +4H^\dag H +K_r^2\right)
\nonumber
\\
 &\!\!\!=\!\!\!&
    16\lambda_1 f_S^2  (\sqrt{2} \bar{D}_r +\sqrt{2}\bar{J}_r 
        +\bar{K}_r)^2
      +16 \lambda_2 f_S (\bar{D}_r^2 +\bar{J}_r^2 +\bar{K}_r^2)
\nonumber
\\
  &&
       +32 \lambda_2 f_S
   (\bar{A}\bar{A}^\dag + \bar{B}\bar{B}^\dag +\bar{F}\bar{F}^\dag
        +\bar{H}^\dag \bar{H} +   \bar{E}\bar{E}^\dag) .
\eea
By transforming the basis into the mass eigenstates
via the orthogonal matrix,
\bea
 && \left(\begin{array}{c}
    I_1 \\
    I_2 \\
    I_3 \\
    \end{array}\right)
    \equiv
   {1\over \sqrt{10}} 
     \left(\begin{array}{ccc}
     1 & 1 & -2\sqrt{2} \\
     \sqrt{5} & -\sqrt{5} & 0 \\
     2 & 2 & \sqrt{2} \\
     \end{array}\right) 
   \left(\begin{array}{c}
    \bar{D_r} \\
    \bar{J_r} \\
    \bar{K_r} \\
    \end{array}\right) ,
\eea 
the mass terms are diagonalized as
\bea
  V_{R^2} =
\textrm{${1\over 2}$} M_{I_1}^2 I_1^2
  +\textrm{${1\over 2}$} M_{I_2}^2 I_2^2
  +\textrm{${1\over 2}$} M_{I_3}^2 I_3^2 
  +32 \lambda_2 f_S
   (\bar{A}\bar{A}^\dag + \bar{B}\bar{B}^\dag +\bar{F}\bar{F}^\dag
        +\bar{H}^\dag \bar{H} +   \bar{E}\bar{E}^\dag) .
 \label{vr2}
\eea
Here the masses squared are given by
\bea
  M_{I_1}^2 = M_{I_2}^2 = 32\lambda_2 f_S^2 , \qquad
  M_{I_3}^2 = 32 (5\lambda_1 +\lambda_2) f_S^2 .
  \label{masses}
\eea

\subsubsection*{Scalar-gauge interactions}

Now we derive the couplings 
relevant to the Higgs mass correction through gauge boson loop.
The gauge interaction is included in Eq.~(\ref{covs}).
In the breaking $\textrm{SU}(5)\to\left[\textrm{SU(2)}\times\textrm{U(1)}\right]^2$,
 only $W_{\mu\,i}$ and $B_{\mu\,i}$
 ($i=1,2$)
have masses below 10~TeV.
As a low energy effective theory, the coupling of
$W_{\mu\,i}$ and $B_{\mu\,i}$ with scalar fields is needed.
Under the group $\left[\textrm{SU(2)}\times\textrm{U(1)}\right]^2$,
the charges of $S$ can be assigned as
\bea
 && Q_1^a = \left(\begin{array}{ccc}
      \tau^a  & &  \\
         & 0 & \\ 
         &  & {\bf 0}_2 \\
      \end{array} \right) , \qquad
  Y_1 ={1\over 10}\left(\begin{array}{ccc}
        3 \cdot {\bf 1}_2 & & \\
          &-2 & \\ 
          && -2 \cdot {\bf 1}_2 \\ 
        \end{array}\right) ,
\\
&& Q_2^a = \left(\begin{array}{ccc}
      {\bf 0}_2  & &  \\ 
         & 0 & \\ 
         &  & -\tau^a{}^* \\
      \end{array} \right) , \qquad
  Y_2 ={1\over 10}\left(\begin{array}{ccc}
        2\cdot {\bf 1}_2 & & \\
           &2 & \\ 
          && -3\cdot {\bf 1}_2 \\ 
        \end{array}\right) .
\eea
To identify the couplings, it is convenient to
write down the second and third terms in
the covariant derivative (\ref{covs}) as
\bea
  &&  -i g A^a T^a S -i g A^a ST^a{}^T 
\nonumber
\\
 &\!\!\!=\!\!\!&   
    -ig(W_1^a-W_2^a) 
    f_S\left(\begin{array}{ccc}
       &&\tau^a\\
       &0&\\
       \tau^a{}^*&&\\
       \end{array} \right)
     -i g'(B_1-B_2)
      {1\over 10}
    f_S\left(\begin{array}{ccc}
       &&{\bf 1}_2\\
       &-4&\\
       {\bf 1}_2&&\\
       \end{array} \right)
\nonumber
\\
 &&
     +g\left(\begin{array}{ccc}
       W_1^a(\tau^a \phi_- +\phi_- \tau^a{}^*) 
                              & W_1^a\tau^a h_- 
               &W_1^a \tau^a \chi_- -W_2^a \chi_- \tau^a \\
       W_1^a h_-^T \tau^a{}^*  & 0 & -W_2^a h_+^\dag \tau^a  \\
       W_1^a \chi_-^T\tau^a{}^* 
       -W_2^a \tau^a{}^* \chi_-^T  
    & -W_2^a \tau^a{}^* h_+^*& 
    -W_2^a (\tau^a{}^* \phi_+^\dag +\phi_+^\dag \tau^a)\\
       \end{array}\right)
\nonumber
\\
 &&
     +{g'\over 10}\left(\begin{array}{ccc}
       (6B_1+4B_2) \phi_- & (B_1 +4B_2) h_- & (B_1-B_2) \chi_- \\
       (B_1+4B_2)  h_-^T & -4(B_1-B_2)(K_i-iK_r) & -(4B_1+B_2) h_+^\dag \\
       (B_1-B_2)  \chi_-^T & -(4B_1+B_2) h_+^* & -(4B_1+6B_2) \phi_+^\dag \\
       \end{array}\right)       ,
    \label{fset}
\eea
where
$\phi_\pm = \phi \pm i \Phi$,
$h_\pm = h \pm iH$,
$\chi_- = \chi -i X$
and
$g' =\sqrt{5/3} g$.
The dominant gauge interactions relevant to 
the scalar mass correction
are symbolically given by $A^2$, $A^2 N^2$ and $A^2R$ terms
at lower order of $R$.
We obtain each term as follows:
The $A^2$ term is
\bea
   \textrm{${1\over 8}$} \textrm{Tr}
     \left[(D_\mu S)(D^\mu S)^\dag \right]_{A^2}
  =
   \textrm{${1\over 8}$} g^2 f_S^2 (W_1^a - W_2^a)^2
 +\textrm{${1\over 40}$} g'{}^2 f_S^2 (B_1-B_2)^2 .
\eea
The $A^2N^2$ term is
\bea
&& \textrm{${1\over 8}$} \textrm{Tr}
     \left[(D_\mu S)(D^\mu S)^\dag \right]_{A^2N^2}
\nonumber
\\
  &=\!\!\!&
  \textrm{${1\over 16}$} g^2 (W_1^a {}^2 + W_2^a {}^2)
  \, \textrm{Tr} \left[\phi\phi^\dag \right]
   +\textrm{${1\over 4}$}g^2 (W_1^a W_1^b + W_2^a W_2^b) \,
 \textrm{Tr} \left[\tau^a \phi \tau^b{}^* \phi^\dag \right]
\nonumber
\\
  && +\textrm{${1\over 16}$} g^2 (W_1^a{}^2 + W_2^a {}^2)
  h^\dag h
\nonumber
\\
 &&+\textrm{${1\over 16}$} g^2 (W_1^a{}^2 + W_2^a {}^2)
  \,\textrm{Tr} \left[\chi^2 \right]
  -\textrm{${1\over 2}$}g^2 W_1^a W_2^b \,\textrm{Tr}
  \left[\tau^a \chi \tau^b \chi \right]
\nonumber
\\
  && +\textrm{${1\over 200}$}g'{}^2 (13B_1^2 +24B_1B_2 +13B_2^2) \,
  \textrm{Tr} \left[\phi\phi^\dag \right]
\nonumber
\\
  && +\textrm{${1\over 400}$}g'{}^2 (17B_1^2 +16B_1 B_2 +17B_2^2) h^\dag h
\nonumber
\\
  && +\textrm{${1\over 400}$}g'{}^2 (B_1-B_2)^2 \, \textrm{Tr}
  \left[ \chi^2\right] 
  +\textrm{${1\over 50}$} g'{}^2 (B_1 -B_2)^2 K_i^2
\nonumber
\\
  && + \textrm{${1\over 20}$}gg' (6B_1 +4B_2) W_1^a\,
  \textrm{Tr} \left[ \tau^a
   \phi\phi^\dag \right]
   + \textrm{${1\over 20}$}gg' (4B_1 +6B_2) W_2^a\,
  \textrm{Tr} \left[ \tau^a
   \phi\phi^\dag \right]
\nonumber
\\
  &&+\textrm{${1\over 20}$}gg' (B_1-B_2) (W_1^a-W_2^a)\,
  \textrm{Tr} \left[ \tau^a \chi^2 
  \right]
\nonumber
\\
  && +\textrm{${1\over 20}$}gg' (B_1 +4B_2)W_1^a
  h^\dag \tau^a h
   +\textrm{${1\over 20}$}gg' (4B_1 +B_2)W_2^a
  h^\dag \tau^a h
 .
  \label{va2n2}
\eea
The $A^2 R$ term is
\bea
 && \textrm{${1\over 8}$} \textrm{Tr}
     \left[(D_\mu S)(D^\mu S)^\dag \right]_{A^2 R}
\nonumber
\\
 &\!\!\!=\!\!\!&     
 \textrm{${1\over 8}$}g^2 f_S (W_1^a -W_2^a)^2 \,\textrm{Tr} \left[X\right] 
  + \textrm{${1\over 20}$}gg' f_S (B_1-B_2) (W_1^a-W_2^a) \, \textrm{Tr}
  \left[ \tau^a X\right]
\nonumber
\\
  &&
  +\textrm{${1\over 200}$} g'{}^2 f_S (B_1-B_2)^2 
    \, \textrm{Tr}\left[X\right]
    +\textrm{${1\over 25}$} g'{}^2 f_S (B_1-B_2)^2 K_r .
    \label{va2r}
\eea
The equations~(\ref{vn2r}), (\ref{vr2}), (\ref{kins}), 
(\ref{va2n2}) and
(\ref{va2r}) are all the vertices needed for
the scalar mass corrections via gauge boson loop
at lower order of $R$.
From Eq.~(\ref{va2n2}),
the tree vertex 
is given by the contact term
\bea
   \textrm{${1\over 8}$}
     \textrm{Tr} \left[(D_\mu S)(D^\mu S)^\dag\right]_{A^2N^2
  \, *}
  &\!\!\! =\!\!\!&
    \textrm{${1\over 16}$}
      g^2 (W_1^a{}^2 + W_2^a {}^2)( \textrm{Tr}\left[\phi\phi^\dag\right]
        +h^\dag h+ \textrm{Tr}\left[\chi^2\right] )
\nonumber
\\
&& +\textrm{${1\over 200}$} g'{}^2 
  (13 B_1^2 +24 B_1 B_2 +13 B_2^2 )\textrm{Tr} \left[\phi\phi^\dag\right]
\nonumber
\\
 && +\textrm{${1\over 400}$} g'{}^2
    (17 B_1^2 + 16 B_1 B_2 +17 B_2^2) h^\dag h
\nonumber
\\
  && +\textrm{${1\over 400}$} g'{}^2
    (B_1 - B_2)^2 \textrm{Tr}\left[\chi^2\right]
      +\textrm{${1\over 50}$} g'{}^2 (B_1-B_2)^2 K_i^2  .
    \label{contact}
\eea
The effective $A^2 N^2$ vertex is made through integration of 
the heavy field $R$ with the vertices $V_{N^2 R}$ and $V_{A^2 R}$. 
From Eq.~(\ref{vn2r}), the corresponding Lagrangian term
for $N^2 R$ vertex is 
\bea
- V_{N^2 R \, *} &\!\!\! =\!\!\!& 
  -4f_S \lambda_1 (2\textrm{Tr}(X) + K_r)
    (2\textrm{Tr} (\phi\phi^\dag) +2\textrm{Tr} (\chi^2) +K_i^2 + 4h^\dag h)
\nonumber
\\
  && -4f_S \lambda_2 K_r (2h^\dag h +K_i^2)
    -4f_S \lambda_2 \textrm{Tr}(X) \textrm{Tr} \left[
      \phi\phi^\dag + hh^\dag + \chi^2\right] ,
      \label{kri}
\eea
where we have used a decomposition $X={1\over 2} \textrm{Tr} (X) {\bf 1}_2
  + \left(X-{1\over 2}\textrm{Tr} (X) {\bf 1}_2\right)$. 
From Eq.~(\ref{kri}), it is seen that the heavy fields 
$\textrm{Tr}(X)=2(I_1+I_3)/\sqrt{5}$ and
$K_r=2(-2I_1+I_3)/\sqrt{5}$ make contributions to the effective vertex.
The contributing Lagrangian term of the $A^2 R$ vertex is
\bea
&&
 \textrm{${1\over 8}$} \textrm{Tr}\left[
   (D_\mu S)(D^\mu S)^\dag \right]_{A^2R\, *}
\nonumber
\\
&=\!\!\!&
   \textrm{${1\over 8}$} g^2 f_S (W_1^a -W_2^a)^2 \textrm{Tr}\left(X\right)
 +  \textrm{${1\over 200}$} g'{}^2 f_S (B_1-B_2)^2 
  \left(\textrm{Tr}\left(X\right)
    +8K_r\right) .
     \label{wwr2}
\eea
From Eqs.~(\ref{masses}), (\ref{kri}) and (\ref{wwr2}),
the $A^2 N^2$ term by integral of the field $R$ becomes
\bea
    -V_{A^2 N^2 \, \textrm{\scriptsize Int}}
     &\!\!\!=\!\!\!&-\textrm{${1\over 16}$} g^2 (W_1^a-W_2^a)^2 \left[
         \textrm{Tr}(\phi\phi^\dag) +\textrm{Tr} (\chi^2) +h^\dag h \right]    
\nonumber
\\
  && -\textrm{${1\over 16}$}
    g'{}^2 (B_1-B_2)^2 \left[\textrm{Tr} (\phi\phi^\dag)
      +\textrm{Tr} (\chi^2)\right]
\nonumber
\\
  && -\textrm{${1\over 50}$}
    g'{}^2(B_1-B_2)^2 K_i^2 
    -\textrm{${17\over 400}$} g'{}^2 (B_1-B_2)^2 h^\dag h .
\label{rinte}
\eea
Therefore we obtain the Lagrangian term of effective vertex 
$V_{A^2 N^2 \,\textrm{\scriptsize Eff}}$ as
\bea
 -V_{A^2 N^2 \,\textrm{\scriptsize Eff}}
  &\!\!\!=\!\!\!&
   \textrm{${1\over 8}$}
     \textrm{Tr} \left[(D_\mu S)(D^\mu S)^\dag\right]_{A^2N^2
  \, *}
 +(-V_{A^2 N^2\, \textrm{\scriptsize Int}})
\nonumber
\\  
  &\!\!\! =\!\!\!&
      (\textrm{${1\over 4}$}g^2 W_1^a W_2^a 
 +\textrm{${1\over 8}$}g'{}^2 B_1 B_2) h^\dag h
 +\textrm{${1\over 4}$}
      g^2 W_1^a W_2^a ( \textrm{Tr}\left[\phi\phi^\dag\right]
        +\textrm{Tr}\left[\chi^2\right] )
\nonumber
\\
&& +\textrm{${1\over 400}$} g'{}^2 
  (B_1^2 + B_2^2 + 49 B_1 B_2 )\textrm{Tr} \left[\phi\phi^\dag\right]
   -\textrm{${3\over 50}$}
    g'{}^2 (B_1-B_2)^2 \textrm{Tr} (\chi^2) .
    \label{main1}
\eea
As expected,
$h^\dag h$ does not have $W_1^a{}^2$, $W_2^a{}^2$, $B_1^2$ and $B_2^2$ terms.
Thus for the mass of the Higgs field $h$, 
there is no quadratic divergence from gauge boson loop.
After log divergence is regularized, the Higgs boson mass squared changes the value with quantum loop effects. The value itself depends on its couplings with the gauge bosons $W_1$, $W_2$, $B_1$ and $B_2$. 
On the other hand, $\textrm{Tr}(\phi\phi^\dag)$, $\textrm{Tr}(\chi^2)$
have interacting terms with $B_1^2$ and $B_2^2$. Due to these U(1) factors, the masses of $\phi$ and $\chi$ receive quadratic divergence. The coupling for $\chi$ has a negative sign. If theory is arranged such that the mass scale of heavy fields is stabilized, it should be traced carefully whether corrections to scalar masses squared have a positive sign.

\section{Operators in breaking by boundary conditions \label{bclrhs}}

We have found the effective couplings in the model with gauge symmetry broken by the vacuum expectation value of the complex symmetric tensor $S$.
Instead of such a vacuum expectation value,
we examine the case in which 
the two breaking sectors in higher energy scales are spatially separated.
At $y=0$, SU(5) is broken to 
$[\textrm{SU(2)}\times\textrm{U(1)}]^2$ and
at $y=\pi R$, SU(5) is broken to SO(5).
Here $y$ is the coordinate
of the fifth dimension and $R$ is the compactification radius. 
We assume that the five-dimensional spacetime is flat.

\subsection{Consistency of boundary conditions \label{consis}}

The gauge symmetry breaking of SU(5) to SO(5) by boundary conditions at $y=\pi R$ is
obtained by imposing Neumann boundary conditions at $y=\pi R$
only on fields for the generators in Eq.~(\ref{so5l}).
There are no scalar fields required for this gauge symmetry breaking.
On the other hand, extra-dimensional components of bulk gauge bosons can
have nonzero values as four-dimensional scalar fields.
In such a situation, the consistency of boundary conditions with
gauge transformations is nontrivial.
In this subsection, we give two examples of inconsistent boundary conditions
for distinct patterns of gauge symmetry breaking.
Then we check the consistency of the gauge symmetry breaking of SU(5) to SO(5).

\subsubsection*{A model of $\textrm{SU}(N) \to \textrm{SU}(N_1) \times \textrm{U}(1)$}
We consider a gauge symmetry breaking of SU($N$), 
\bea
  \textrm{SU}(N)\to
  \textrm{SU}(N_1)\times\textrm{SU}(N-N_1)\times \textrm{U}(1)
  \to
  \textrm{SU}(N_1)\times \textrm{U}(1) .
\eea
Here we assume the following boundary conditions \cite{Sakai:2006qi}
as shown in Table~\ref{tabselect}:
Neumann (N) for $A_\mu^a(x,y)$, Dirichlet (D) for $A_y^a(x,y)$,
D for $A_\mu^{\hat{a}}(x,y)$ and N for $A_y^{\hat{a}}(x,y)$ at $y=0$;
D for $A_\mu^a(x,y)$, N for $A_y^a(x,y)$,
N for $A_\mu^{\hat{a}}(x,y)$ and D for $A_y^{\hat{a}}(x,y)$ at $y=\pi R$,
where $a$ and $\hat{a}$ indicate 
the indices of SU($N-N_1$) and 
$\textrm{SU}(N)/[\textrm{SU}(N_1)\times\textrm{SU}(N-N_1)\times \textrm{U}(1)]$,
respectively. We omit the boundary conditions for
the groups $\textrm{SU}(N_1)$ and U(1).
The boundary conditions for gauge transformation functions are
the same as that of the four-dimensional gauge bosons:
N for $\epsilon^a(x,y)$ and D for $\epsilon^{\hat{a}}(x,y)$ 
at $y=0$; D for $\epsilon^a(x,y)$ and N for $\epsilon^{\hat{a}}(x,y)$
at $y=\pi R$.
\begin{table}[htb]
\begin{center}
\caption{The boundary conditions for
$\textrm{SU}(N)\to \textrm{SU}(N_1)\times \textrm{U}(1)$.} \label{tabselect}
\begin{tabular}{cccccc|cccccc}
\hline\hline
\multicolumn{6}{c}{$y=0$} & \multicolumn{6}{|c}{$y=\pi R$} \\ \hline
$A_\mu^a$ & $A_y^a$ & $A_\mu^{\hat{a}}$  & $A_y^{\hat{a}}$ & 
$\epsilon^a$ & $\epsilon^{\hat{a}}$ &
$A_\mu^a$ & $A_y^a$ & $A_\mu^{\hat{a}}$  & $A_y^{\hat{a}}$ &
$\epsilon^a$ & $\epsilon^{\hat{a}}$ \\ \hline
N & D & D & N & N & D &
D & N & N & D & D & N  \\ \hline
\end{tabular}
\end{center}
\end{table}

For each color, the gauge transformation laws are given by
\bea
   \delta A_M^a  &\!\!\!=\!\!\!& \partial_M \epsilon^a
    +gf^{abc} A_M^b \epsilon^c
    +gf^{a\hat{b}\hat{c}} A_{M}^{\hat{b}} \epsilon^{\hat{c}},
 \label{gauge1} 
\\
   \delta A_M^{\hat{a}}  &\!\!\!=\!\!\!& \partial_M \epsilon^{\hat{a}}
    +gf^{\hat{a}b\hat{c}} A_M^b \epsilon^{\hat{c}}
    +gf^{\hat{a}\hat{b}c} A_M^{\hat{b}} \epsilon^{c}, 
   \label{gauge2}
\eea
where $M=(\mu,y)$.
The boundary conditions of the left-hand and right-hand sides 
for the gauge transformations (\ref{gauge1}) and (\ref{gauge2}) are
tabulated in Table~\ref{tablrbc}.
\begin{table}[htb]
\begin{center}
\caption{The boundary conditions for the terms of the transformation laws.} \label{tablrbc}
\begin{tabular}{c|c|c|c|c}
\hline\hline
 & \multicolumn{2}{|c}{$y=0$} 
 & \multicolumn{2}{|c}{$y=\pi R$} \\  \cline{2-5}
 & LHS & RHS & LHS & RHS \\ \hline
 $A_\mu^a$ & N & N $+$ NN $+$ DD
   & D & D $+$ DD $+$ NN \\   
 $A_y^a$ & D & D $+$ DN $+$ ND
   & N & N $+$ ND $+$ DN \\
 $A_\mu^{\hat{a}}$ & D & D $+$ ND $+$ DN
   & N & N $+$ DN $+$ ND \\
 $A_y^{\hat{a}}$ & N & N $+$ DD $+$ NN
   & D & D $+$ NN $+$ DD \\ \hline
\end{tabular}
\end{center}
\end{table}
From Table~\ref{tablrbc},
the boundary conditions at $y=0$ 
are consistent with the gauge transformations.
It is seen that 
the other boundary conditions are inconsistent because for example
the NN term for $A_\mu^a$ does not obey Dirichlet condition,
$(A_\mu^{\hat{b}}\epsilon^{\hat{c}})|_{y=\pi R}\neq 0$ 
and the ND term for $A_y^a$ does not obey Neumann condition,
$\partial_y(A_y^b \epsilon^c)|_{y=\pi R}=
(A_y^b \partial_y \epsilon^c)|_{y=\pi R} \neq 0$.

\subsubsection*{A left-right symmetric model}

As another example, we discuss a left-right symmetric model with
the gauge group $\textrm{SU}(2)_L\times\textrm{SU}(2)_R\times \textrm{U}(1)$
broken as
\bea
&&   \textrm{SU}(2)_L\times \textrm{SU}(2)_R
    \times \textrm{U}(1) \to 
    \textrm{SU}(2)_D \times \textrm{U}(1) ,~~\textrm{at} ~ y=0 ,
\\
 &&  \textrm{SU}(2)_L\times \textrm{SU}(2)_R
    \times \textrm{U}(1) \to 
    \textrm{SU(2)}_L \times \textrm{U}(1)_Y ,~~\textrm{at} ~y=\pi R. 
\eea
The gauge bosons of SU(2)${}_L$, SU(2)${}_R$ and
U(1) are denoted as $A_{L \,\mu}^a$, $A_{R\,\mu}^a$ and $B_\mu$, respectively.
The gauge coupling constants are identical for SU(2)${}_L$ and SU(2)${}_R$ and
it is denoted as $g$. The gauge coupling constant of U(1) is denoted as $g'$.
The boundary conditions are given in Table~\ref{tablr}.
\begin{table}[htb]
\begin{center}
\caption{The boundary conditions for the left-right symmetry model.} \label{tablr}
\begin{tabular}{ccc|ccc}
\hline\hline
\multicolumn{3}{c}{$y=0$} & \multicolumn{3}{|c}{$y=\pi R$} \\ \hline
$A_{L+R \,\mu}^a$ & $A_{L-R\, \mu}^a$ & $B_\mu$ &
$A_{L\,\mu}^a$ & $A_{R\,\mu}^{1,2}$, $\bar{A}_{R\,\mu}^{3}$
& $A_{Y\,\mu}$ \\ \hline
N & D & N & N & D & N \\ \hline
\end{tabular}
\end{center}
\end{table} 

\noindent
Here we have defined 
\bea
  \left(\begin{array}{c}
    A_{L+R}^a \\
    A_{L-R}^a \\ 
     \end{array}\right)
     = \textrm{${1\over \sqrt{2}}$} \left(\begin{array}{cc}
         1 & 1 \\
         1 & -1 \\
         \end{array}\right)
         \left(\begin{array}{c}
           A_L^a \\
           A_R^a \\
           \end{array}\right)
        ,~
 \left(\begin{array}{c}
   \bar{A}_{R}^3 \\
   A_{Y} \\
   \end{array}\right)
   ={1\over \sqrt{g^2 + g'{}^2}}
     \left(\begin{array}{cc}
        g & -g' \\
        g' & g \\
        \end{array}\right)
        \left(\begin{array}{c}
        A_R^3 \\
        B \\
        \end{array}\right) ,
\eea
where the vector index $M$ has been omitted.
The gauge transformation functions $\epsilon_L^a$, $\epsilon_R^a$ and
$\epsilon_B$ have the same boundary conditions as
the gauge bosons with the definition
\bea
 \left(\begin{array}{c}
    \epsilon_{L+R}^a \\
    \epsilon_{L-R}^a \\ 
     \end{array}\right)
     = {1\over \sqrt{2}} \left(\begin{array}{cc}
         1 & 1 \\
         1 & -1 \\
         \end{array}\right)
         \left(\begin{array}{c}
           \epsilon_L^a \\
           \epsilon_R^a \\
           \end{array}\right), ~
 \left(\begin{array}{c}
   \bar{\epsilon}_R^3 \\
   \epsilon_Y \\
   \end{array}\right)
   ={1\over \sqrt{g^2 + g'{}^2}}
     \left(\begin{array}{cc}
        g & -g' \\
        g' & g \\
        \end{array}\right)
        \left(\begin{array}{c}
        \epsilon_R^3 \\
        \epsilon_B \\
        \end{array}\right) . 
\eea
The extra-dimensional components have 
the opposite boundary conditions to the four-dimensional components.
The gauge transformation laws are written in terms of N and D fields at $y=0$ as
\bea
   \delta A_{L+R\,M}^a  &\!\!\!=\!\!\!& \partial_M \epsilon_{L+R}^a
    +gf^{abc} A_{L+R\,M}^b \epsilon_{L+R}^c
    +gf^{abc} A_{L-R\,M}^{b} \epsilon_{L-R}^{c},  
  \label{bclr1} 
\\
   \delta A_{L-R\,M}^a  &\!\!\!=\!\!\!& \partial_M \epsilon_{L-R}^a
    +gf^{abc} A_{L+R\,M}^b \epsilon_{L-R}^c
    +gf^{abc} A_{L-R\,M}^{b} \epsilon_{L+R}^{c}, 
  \label{bc2lr}
\\
   \delta B_M  &\!\!\!=\!\!\!& \partial_M \epsilon_B ,
 \label{bc3lr}
\eea
and in terms of N and D fields at $y=\pi R$  as
\bea
   \delta A_{L\,M}^a  &\!\!\!=\!\!\!& \partial_M \epsilon_{L}^a
    +gf^{abc} A_{L\,M}^b \epsilon_{L}^c, 
\\
   \delta A_{R\,M}^1  &\!\!\!=\!\!\!& \partial_M \epsilon_{R}^1
    +\textrm{${g^2\over \sqrt{g^2 +g'{}^2}}$}(A_{R\,M}^2 \bar{\epsilon}_{R}^3
    -\bar{A}_{R\,M}^3 \epsilon_R^2)
  -\textrm{${gg'\over \sqrt{g^2 +g'{}^2}}$}(A_{R\,M}^2 \epsilon_Y
     -A_{Y\,M} \epsilon_R^2)  ,
\\
   \delta A_{R\,M}^2  &\!\!\!=\!\!\!& \partial_M \epsilon_{R}^2
    -\textrm{${g^2\over \sqrt{g^2 +g'{}^2}}$}(A_{R\,M}^1 \bar{\epsilon}_{R}^3
    -\bar{A}_{R\,M}^3 \epsilon_R^1)
  +\textrm{${gg'\over \sqrt{g^2 +g'{}^2}}$}(A_{R\,M}^1 \epsilon_Y
     -A_{Y\,M} \epsilon_R^1)   ,  
\\
   \delta \bar{A}_{R\,M}^3 &\!\!\!=\!\!\!& \partial_M \bar{\epsilon}_{R}^3
    +\textrm{${g^2\over \sqrt{g^2 +g'{}^2}}$}(A_{R\,M}^1 \epsilon_{R}^2
    -A_{R\,M}^2 \epsilon_{R}^1),
\\
   \delta A_{Y\,M} &\!\!\!=\!\!\!& \partial_M \epsilon_{Y}
    +\textrm{${gg'\over \sqrt{g^2 +g'{}^2}}$}(A_{R\,M}^1 \epsilon_{R}^2
    -A_{R\,M}^2 \epsilon_{R}^1).
\eea
The boundary conditions of the left- and right-hand sides are shown
in Table~\ref{bctlr}.
\begin{table}[htb]
\begin{center}
\caption{The boundary conditions for the terms of the transformation laws.}
\label{bctlr}
\begin{tabular}{ccc}
\hline\hline 
& \multicolumn{2}{|c}{$y=0$} \\ \cline{2-3}
& \multicolumn{1}{|c}{LHS} & \multicolumn{1}{|c}{RHS} \\ \hline
$A_{L+R\,\mu}^a$ & \multicolumn{1}{|c}{N} & \multicolumn{1}{|c}{N $+$ NN $+$ DD} \\
$A_{L+R\, y}^a$ & \multicolumn{1}{|c}{D} & \multicolumn{1}{|c}{D $+$ DN $+$ ND} \\
$A_{L-R\,\mu}^a$ & \multicolumn{1}{|c}{D} & \multicolumn{1}{|c}{D $+$ ND $+$ DN} \\
$A_{L-R\, y}^a$ & \multicolumn{1}{|c}{N} & \multicolumn{1}{|c}{N $+$ DD $+$ NN} \\
$B_\mu$ & \multicolumn{1}{|c}{N} & \multicolumn{1}{|c}{N} \\
$B_y$ & \multicolumn{1}{|c}{D} & \multicolumn{1}{|c}{D} \\
\hline \\
&& \\
\end{tabular} 
~~~
\begin{tabular}{c|c|c}
\hline\hline 
& \multicolumn{2}{|c}{$y=\pi R$} \\ \cline{2-3}
& LHS & RHS \\ \hline
$A_{L\,\mu}^a$ & N & N $+$ NN  \\
$A_{L\, y}^a$ & D & D $+$ DN  \\
$A_{R\,\mu}^{1,2}$ & D & D $+$ DD $+$ DN $+$ ND \\
$A_{R\, y}^{1,2}$ & N & N $+$ ND $+$ NN $+$ DD \\
$\bar{A}_{R\,\mu}^3$ & D & D $+$ DD \\
$\bar{A}_{R\, y}^3$ & N & N $+$ ND \\
$A_{Y\, \mu}$ & N & N $+$ DD \\
$A_{Y\, y}$ & D & D $+$ ND \\
\hline
\end{tabular}
\end{center}
\end{table}

\noindent
From Table~\ref{bctlr},
the boundary conditions at $y=0$ 
are consistent with the gauge transformations.
The inconsistency arises from the ND terms of $A_{R\, y}^{1,2}$ and
$\bar{A}_{R\, y}^3$ because the ND term does not obey Neumann condition.

\subsubsection*{Consistent boundary conditions for SU(5)$\to$ SU(2)$\times$U(1)}

We have seen the two examples which have the inconsistency 
between gauge transformation laws
and boundary conditions.
It is needed to examine the consistency of 
possible boundary conditions to yield gauge symmetry breaking
$\textrm{SU}(5)\to \left[ \textrm{SU}(2)\times \textrm{U}(1)\right]^2$
at $y=0$ and
$\textrm{SU}(5)\to \textrm{SO}(5)$ at $y=\pi R$. 
In order to realize the symmetry breaking above,
we assign Neumann condition for the gauge bosons of the generators 
$T_1$, $T_2$, $T_3$, $T_8$, $T_{15}$, $T_{22}$,
$T_{23}$, $T_{24}$ at $y=0$ and  
$T_{\bar{1}},\cdots, T_{\bar{10}}$ given in Eq.~(\ref{so5l})
at $y=\pi R$ and Dirichlet condition for the other gauge bosons.
Only the fields with Neumann condition at both boundaries have zero modes.
The generators of zero modes are
\bea
  T_{\overline{1}} , T_{\overline{2}},
  T_{\overline{9}}, T_{\overline{10}} .
 \label{bcta}
\eea
These generators form SU(2)$\times$U(1) algebra.
The gauge transformation laws are written as
\bea
  \delta A_M^{a} &\!\!\!=\!\!\!&\partial_M \epsilon^{a}
    + gf^{abc} A_M^{b} \epsilon^{c}
  + gf^{a\hat{b}\hat{c}} A_M^{\hat{b}} 
  \epsilon^{\hat{c}} ,
\\
   \delta A_M^{\hat{a}} &\!\!\!=\!\!\!&
  \partial_M \epsilon^{\hat{a}}
    + gf^{\hat{a}b\hat{c}} A_M^{b}\epsilon^{\hat{c}}
  + gf^{\hat{a}b\hat{c}} A_M^{b} 
  \epsilon^{\hat{c}} ,
\eea
where $a$ and $\hat{a}$ indicate the generators of the subgroup and
the coset, respectively.
At $y=0$, $a$ and $\hat{a}$ represent the indices for
$\left[ \textrm{SU}(2)\times \textrm{U}(1)\right]^2$
and $\textrm{SU}(5)/\left[ \textrm{SU}(2)\times \textrm{U}(1)\right]^2$, respectively.
At $y=\pi R$, $a$ and $\hat{a}$ represent the indices for
SO(5) and SU(5)/SO(5), respectively.
The boundary conditions at $y=0$ and $y=\pi R$ are collectively written 
with the $a$ and $\hat{a}$. 
The left- and right-hand sides in the transformation laws have
the boundary conditions shown in Table~\ref{bct55}.
\begin{table}[htb]
\begin{center}
\caption{The boundary conditions for 
$\textrm{SU}(5)\to\textrm{SU}(2)\times\textrm{U}(1)$.} \label{bct55}
\begin{tabular}{c|c|c}
\hline\hline
 & LHS & RHS \\ \hline
$A_\mu^a$ & N & N $+$ NN $+$ DD \\
$A_y^a$ & D & D $+$ DN $+$ ND \\
$A_\mu^{\hat{a}}$ & D & D $+$ DN $+$ ND \\
$A_y^{\hat{a}}$ & N & N $+$ NN $+$ DD \\ \hline
\end{tabular}
\end{center}
\end{table}
From Table~\ref{bct55}, it is found that all the boundary conditions are 
consistent with the gauge transformations.
The result of the consistency in this case is also seen from the fact that 
Neumann and Dirichlet conditions imposed at each boundary 
can be assigned by
the automorphism of orbifolds.

\subsection{Interactions of scalar fields with gauge bosons}

In the gauge symmetry breaking by boundary conditions,
there exist zero-mode scalar fields.
From the boundary conditions given above Eq.~(\ref{bcta}),
$A_y$ has Neumann condition 
for the 16 generators $T_4, \cdots, T_7$, $T_9, \cdots, T_{14}$,
$T_{16}, \cdots, T_{21}$ at $y=0$ and
the 14 generators $T_{\overline{11}}, \cdots, T_{\overline{20}}$,
$T_9, T_{10}, T_{18}, T_{19}$ at $y=\pi R$.
The common 10 generators $T_{\overline{13}}, \cdots, T_{\overline{18}}$,
$T_9, T_{10}, T_{18}, T_{19}$ correspond to zero modes of scalar fields.

To identify the couplings of the scalar fields with gauge bosons, we focus on
the Lagrangian term
\bea
    -\textrm{${1\over 2}$} g^2 \, f^{abc} f^{dec} A_\mu^a A_y^b A^{\mu d} A_y^e .
    \label{mumuyy}
\eea
Keeping the zero modes of gauge bosons $A_\mu^{\overline{1}}$,
$A_\mu^{\overline{2}}$, $A_\mu^{\overline{9}}$, $A_\mu^{\overline{10}}$ and
the scalar fields $A_y^{\overline{13}},\cdots, A_y^{\overline{18}}$,
$A_y^9,A_y^{10},A_y^{18}, A_y^{19}$,
Eq.~({\ref{mumuyy}) become
\bea
 &&   -{g^2\over 16}\left({(A_\mu^{\overline{1}})}^2 
  + {(A_\mu^{\overline{2}})}^2\right)
   \left({(A_y^{\overline{13}})}^2 + {(A_y^{\overline{14}})}^2 
  + {(A_y^{\overline{15}})}^2 + {(A_y^{\overline{16}})}^2 
  + 4\lbrace{(A_y^{\overline{17}})}^2 +{(A_y^{\overline{18}})}^2\rbrace \right)
\nonumber
\\
  && -{g^2\over 4}
    \left( {(-\textrm{${1\over 2}$}A_\mu^3 + \textrm{${\sqrt{3}\over 2}$} A_\mu^8)}^2
      +{(-\textrm{${\sqrt{3}\over 3}$} A_\mu^8 
    + \textrm{${\sqrt{6}\over 12}$} A_\mu^{15}
        +\textrm{${\sqrt{10}\over 4}$}A_\mu^{24} )}^2 \right)
        \left({(A_y^{\overline{15}})}^2 
+ {(A_y^{\overline{16}})}^2 \right)    
\nonumber
\\
  && -{g^2\over 4}
   \left( {(\textrm{${1\over 2}$} A_\mu^3 + \textrm{${\sqrt{3}\over 2}$}A_\mu^8)}^2
     +{(-\textrm{${\sqrt{3}\over 3}$} A_\mu^8 
   + \textrm{${\sqrt{6}\over 3}$} A_\mu^{15})}^2\right)
     \left({(A_y^{\overline{13}})}^2
       +{(A_y^{\overline{14}})}^2 \right) 
\nonumber
\\
  && -{g^2\over 4}
  \left( {(-\textrm{${1\over 2}$} A_\mu^3 
    +\textrm{${\sqrt{3}\over 6}$} A_\mu^8
    +\textrm{${\sqrt{6}\over 3}$} A_\mu^{15})}^2
    +{(\textrm{${1\over 2}$} A_\mu^3
    +\textrm{${\sqrt{3}\over 6}$}A_\mu^8
    +\textrm{${\sqrt{6}\over 12}$}A_\mu^{15}
    +\textrm{${\sqrt{10}\over 4}$}A_\mu^{24})}^2
    \right)
\nonumber
\\
 &&\qquad
   \times
    \left({(A_y^{\overline{17}})}^2 
      +{(A_y^{\overline{18}})}^2\right)
\nonumber
\\
  && -{\sqrt{2} g^2 \over 8}
     \left(A_\mu^{\overline{1}} A^{\overline{10}\mu}
       (A_y^{\overline{13}} A_y^{\overline{15}}
        +A_y^{\overline{14}} A_y^{\overline{16}})
     +A_\mu^{\overline{2}}A^{\overline{10}\mu}
       (A_y^{\overline{14}} A_y^{\overline{15}}
        -A_y^{\overline{13}} A_y^{\overline{16}})
      \right) 
\nonumber
\\
  && -{g^2\over 8} \left((A_\mu^{\overline{1}})^2 + (A_\mu^{\overline{2}})^2\right)
    \left((A_y^9)^2 +(A_y^{10})^2 +(A_y^{18})^2 +(A_y^{19})^2\right)
\nonumber
\\
  && -{g^2\over 4} \left( (A_\mu^{\overline{1}})^2 -(A_\mu^{\overline{2}})^2\right)
  \left(A_y^9 A_y^{18} + A_y^{10} A_y^{19}\right) 
  -{g^2\over 2} A_\mu^{\overline{1}}A^{\overline{2}\mu}
    \left( -A_y^9 A_y^{19} + A_y^{10} A_y^{18} \right) .
    \label{expa}
\eea    
Here
\bea
  \left(\begin{array}{c}
      A_\mu^3 \\
      A_\mu^8 \\
      A_\mu^{15} \\
      A_\mu^{24} \\
      \end{array} \right)
     =\textrm{${4\sqrt{2} \over 26+3\sqrt{10}}$}
    \left(\begin{array}{cccc}
       4 & {5\over 4}(1-{\sqrt{10}\over 2}) & {5\over 2}+{3\sqrt{10}\over 4} 
   & {5\sqrt{2}\over 8} -{\sqrt{5}\over 4} \\
       0 & {13\sqrt{3}\over 12}+ {\sqrt{30}\over 8} & 0 &
    {5\sqrt{6}\over 8} +{13\sqrt{15}\over 12} \\
       \sqrt{6} & {5\sqrt{6}\over 3} & -\sqrt{6} & -{\sqrt{30}\over 3} \\
       -\sqrt{10} & {5\over 2}+{\sqrt{10}\over 2} & \sqrt{10} 
  & -{\sqrt{2}\over 2} -{\sqrt{5}\over 2} \\
     \end{array}\right)   
 \left(\begin{array}{c}
     A_\mu^{\overline{9}} \\
     A_\mu^{\overline{10}} \\
     A_\mu^{\overline{19}} \\
     A_\mu^{\overline{20}} \\
     \end{array}\right) .
\eea
On the other hand, keeping the gauge bosons for Neumann condition at $y=0$,
$A_\mu^1, A_\mu^2, A_\mu^3$, $A_\mu^8, A_\mu^{15}$, 
$A_\mu^{22}, A_\mu^{23}, A_\mu^{24}$,
Eq.~(\ref{mumuyy}) becomes
\bea
 &&  -{g^2\over 16}
  \left( (A_\mu^1)^2+(A_\mu^2)^2+(A_\mu^{22})^2+ (A_\mu^{23})^2
     \right) 
     \left((A_y^{\overline{13}})^2 +(A_y^{\overline{14}})^2
       +(A_y^{\overline{15}})^2 + (A_y^{\overline{16}})^2 \right)
\nonumber
\\
  && -{g^2\over 8}\left( (A_\mu^1 -A_\mu^{22})^2
  +(A_\mu^2+ A_\mu^{23})^2 \right) \left( (A_y^{\overline{17}})^2
    +(A_y^{\overline{18}})^2\right)
\nonumber
\\
  && -{g^2\over 2} \left(
   \textrm{${1\over 8}$} (A_\mu^3+\sqrt{3}A_\mu^8)^2
    +\textrm{${1\over 6}$}(A_\mu^8-\sqrt{2}A_\mu^{15})^2 \right)
 \left((A_y^{\overline{13}})^2
  +(A_y^{\overline{14}})^2\right)
\nonumber
\\
  && -{g^2\over 2} \left(
   \textrm{${1\over 8}$} (A_\mu^3 -\sqrt{3}A_\mu^8)^2
  +\textrm{${1\over 6}$}(A_\mu^8 -\textrm{${1\over 2\sqrt{2}}$} A_\mu^{15}
    -\textrm{${\sqrt{30}\over 4}$} A_\mu^{24})^2 \right)
      \left((A_y^{\overline{15}})^2 +(A_y^{\overline{16}})^2 \right)
\nonumber
\\
 && -{g^2\over 16}\left(
    (A_\mu^3 -\textrm{${1\over \sqrt{3}}$}A_\mu^8 
  -\textrm{${4\over \sqrt{6}}$}
       A_\mu^{15})^2
        +(A_\mu^3 + \textrm{${1\over \sqrt{3}}$} A_\mu^8
          +\textrm{${1\over \sqrt{6}}$} A_\mu^{15} 
    + \textrm{${\sqrt{10}\over 2}$} A_\mu^{24})^2
          \right)
       \left( (A_y^{\overline{17}})^2 + (A_y^{\overline{18}})^2\right)
\nonumber
\\
  && -{(3+\sqrt{3})g^2 \over 8} A_\mu^8
  \left(A_\mu^1 (A_y^{\overline{13}}A_y^{\overline{15}}
   +A_y^{\overline{14}}A_y^{\overline{16}})
   -A_\mu^2(A_y^{\overline{13}}A_y^{\overline{16}}
    -A_y^{\overline{14}} A_y^{\overline{15}})\right)
\nonumber
\\
  && -{g^2\over 4\sqrt{3}}(A_\mu^8-\textrm{${5\over 4\sqrt{2}}$} A_\mu^{15}
     -\textrm{${\sqrt{30}\over 8}$}A_\mu^{24})
   \left(A_\mu^{22}(A_y^{\overline{13}}A_y^{\overline{15}}
     +A_y^{\overline{14}}A_y^{\overline{16}})
     +A_\mu^{23} (A_y^{\overline{13}}A_y^{\overline{16}}
     -A_y^{\overline{14}}A_y^{\overline{15}})\right)
\nonumber
\\
 && -{g^2\over 8} \left( (A_\mu^1)^2 + (A_\mu^2)^2
   +(A_\mu^{22})^2 +(A_\mu^{23})^2\right)
     \left( (A_y^9)^2 +(A_y^{10})^2 +(A_y^{18})^2 +(A_y^{19})^2\right)
\nonumber
\\
 && -{g^2\over 8} (A_\mu^3+\textrm{${1\over \sqrt{3}}$} A_\mu^8
   +\textrm{${4\over \sqrt{6}}$} A_\mu^{15})^2
     \left((A_y^9)^2 + (A_y^{10})^2\right)
\nonumber
\\
 && -{g^2\over 8} (A_\mu^3 -\textrm{${1\over \sqrt{3}}$} A_\mu^8
    -\textrm{${1\over \sqrt{6}}$} A_\mu^{15} 
  -\textrm{${\sqrt{10}\over 2}$}A_\mu^{24})^2
     \left((A_\mu^{18})^2 +(A_y^{19})^2\right)
\nonumber
\\
 && +{g^2\over 2} \left((A_\mu^1 A_y^9 +A_\mu^2 A_y^{10})
    (A^{22\mu} A_y^{18} + A^{23\mu} A_y^{19})
  + (A_\mu^1 A_y^{10} -A_\mu^2 A_y^9)
    (A^{22\mu} A_y^{19} -A^{23\mu} A_y^{18}) \right).
\nonumber
\\
    \label{exp2}
\eea
In Eqs.~(\ref{expa}) and (\ref{exp2}),
all the scalar fields are coupled to gauge bosons of the form $A_\mu^a A^{a\mu}$ with the same sign. This means that their scalar fields do not have the couplings given in Eqs.(\ref{basic2}) and (\ref{basic3}). The scalar fields are a component of higher-dimensional gauge bosons. If Kaluza-Klein modes are included infinitely, higher-dimensional gauge symmetry may work as a symmetry principle. However, the collective symmetry breaking mechanism would be a distinct property. In a five-dimensional model, the breaking SU(5)$\to$SO(5) where the Higgs boson resides could be driven  by the vacuum expectation values of scalar fields. For such a breaking, it can be seen that the expected cancellation takes place in a parallel way with the four-dimensional model without relying on a summation over the contributions of Kaluza-Klein modes. Here the collective symmetry breaking mechanism works. In the case of the breaking by
boundary conditions, one may wonder whether the cancellation occurs between zero mode and one of any Kaluza-Klein mode without invoking an infinite number of
Kaluza-Klein modes. The analysis here has been given for fields dependent on the five-dimensional coordinates. It can be seen that the couplings for any Kaluza-Klein mode of gauge bosons are the same sign as that of the zero mode. This means that the expected cancellation does not occur. Even though higher-dimensional gauge symmetry is employed to cancel the quadratic divergence, one would need to take into account contributions such as a three-point coupling with a gauge boson and two scalar fields, which are not relevant to the collective symmetry breaking mechanism.

We have examined interactions of zero-mode gauge bosons
with scalar fields. 
In the five-dimensional setup,
there are no three- and four-point scalar couplings
because of $F_{yy}=0$.
Therefore heavy scalar fields do not contribute to changing 
the scalar-gauge interactions
at tree level.

\section{Properties of symmetry breaking and operators 
in the two models \label{interpret}}

As we have derived in Eqs.~(\ref{main1}), (\ref{expa}) and (\ref{exp2}), 
the operators have different forms depending on
the source of symmetry breaking which is vacuum expectation values or 
boundary conditions. 
In this section, we discuss why the difference of the two models is induced,
by comparing the meaning of
$\textrm{SU(5)}\to [\textrm{SU(2)}\times \textrm{U(1)}]^2$.

In the model with scalars having vacuum expectation values,
the scalar-gauge interactions (\ref{main1}) involve
gauge bosons of 
$[\textrm{SU(2)}\times \textrm{U(1)}]^2$ in breaking  
\bea
   \textrm{SU(5)} \to [\textrm{SU(2)}\times \textrm{U(1)}]^2 .
    \label{double}
\eea
Two of the group $[\textrm{SU(2)}\times \textrm{U(1)}]$ have 
identical gauge coupling constants because they are subgroups of a single group.
We cannot switch off only one of the gauge couplings of two $[\textrm{SU(2)}\times \textrm{U(1)}]$.
Instead we consider the breaking into
a single $[\textrm{SU(2)}\times \textrm{U(1)}]$ as
\bea
  \textrm{SU(5)} \to \textrm{SU(2)}\times \textrm{U(1)} , 
   \label{direct}
\eea
with appropriate choices of scalar potentials independently.
In the breaking into a single $\textrm{SU(2)}\times \textrm{U(1)}$,
there exists global SU(3).
The corresponding Nambu-Goldstone bosons have no potential.
The generation of potential is made when two of $[\textrm{SU(2)}\times \textrm{U(1)}]$
are taken into account as in Eq.~(\ref{double}). 
That the potential requires two of groups leads to
the absence of quadratic divergence.
This type of discussion to consider a single $[\textrm{SU(2)}\times \textrm{U(1)}]$
as a variation was given in Ref.~\cite{Csaki:2008se}.
Because the operation of switch-off is not 
necessarily equivalent
to the direct breaking (\ref{direct}), it is nontrivial whether 
resulting scalar-gauge couplings have the form (\ref{basic2}).
We have checked the emergence of the form (\ref{basic2})
by deriving all the scalar-gauge couplings.

In the model with boundary conditions, the breaking (\ref{direct}),
\bea
  \textrm{SU(5)} \to  \textrm{SU(2)} \times \textrm{U(1)}
    \label{direct2}
\eea
is achieved by boundary conditions consistently
as an overlapping at two boundaries.
However, the breaking (\ref{direct2}) at a single boundary 
is forbidden by consistency with local gauge transformation as follows.
The breaking (\ref{direct2})
 needs Dirichlet condition for diagonal blocks with respect to
gauge bosons in a matrix form.
In Section~\ref{consis},
in a model of $\textrm{SU}(N) \to \textrm{SU}(N_1) \times \textrm{U}(1)$ we have
seen that
D for $A_\mu^a(x,y)$, N for $A_y^a(x,y)$,
N for $A_\mu^{\hat{a}}(x,y)$ and D for $A_y^{\hat{a}}(x,y)$ at a single boundary
(in Section~\ref{consis} at $y=\pi R$) is an inconsistent boundary condition.
In addition, we see
that the boundary condition D for $A_\mu^a(x,y)$, N for $A_y^a(x,y)$,
D for $A_\mu^{\hat{a}}(x,y)$ and N for $A_y^{\hat{a}}(x,y)$
is also not compatible with local gauge transformation.
In the same indication as in Table~\ref{tablrbc},
an inconsistent boundary condition is shown in Table~\ref{tabnod}.
\begin{table}[htb]
\begin{center}
\caption{Inconsistent boundary terms of the transformation laws
with Dirichlet condition for $A_\mu^a(x,y)$ and $A_\mu^{\hat{a}}(x,y)$.} \label{tabnod}
\begin{tabular}{c|c|c}
\hline\hline
 & LHS & RHS \\ \hline
 $A_y^a$ & N & N $+$ ND $+$ ND
    \\ \hline
\end{tabular}
\end{center}
\end{table}
Therefore D for $A_\mu^a(x,y)$ is not allowed.
In other words, as a variation of the breaking 
$\textrm{SU(5)} \to [\textrm{SU(2)}
\times \textrm{U(1)}]^2$ for switching off a single $\textrm{SU(2)}\times \textrm{U(1)}$, the breaking $\textrm{SU(5)}\to \textrm{SU(2)}\times \textrm{U(1)}$
at one boundary does not exist.
In the model with boundary conditions, the generation of potential for
$[\textrm{SU(2)}\times \textrm{U(1)}]^2$ is regarded as the generation of
the potential just for
a subgroup with a single gauge coupling constant.
This suggests that the collective breaking does not work.
On the other hand, the pattern of the symmetry breaking 
group itself is identical
to that of the model with scalars having vacuum expectation values.
Like the model of vacuum expectation values, there might be a possibility 
that the scalar potential could be
 accidentally prevented from having quadratic divergence. 
We have shown that it does not occur by deriving all the relevant 
scalar-gauge operators.

\section{Conclusion \label{conc}}

We have derived all the couplings
of scalar-gauge interactions
relevant to scalar mass corrections
 via gauge boson loop in high energy models.
We have found that while the Higgs fields are protected
with the coupling of the form (\ref{basic}) 
from having quadratic 
divergence, the other fields receive quadratic divergence through
loop of the U(1) gauge boson.
As another aspect, we have considered a possibility of gauge symmetry breaking 
by boundary conditions 
as the first step to utilize dynamical symmetry breaking in a gauge-Higgs unification.
In our assignment for boundary conditions,
the same gauge symmetry breaking as in the vacuum expectation value
is produced in a consistent way with local gauge transformations.
In this gauge symmetry breaking, we have presented the couplings 
between zero-mode gauge bosons and scalar fields.
It has been found that there is a difference of the structure between the operators
in the two models
of vacuum expectation values or boundary conditions.

\vspace{4ex}

\subsubsection*{Acknowledgment}

This work is supported in part by Scientific Grants 
from the Ministry of Education
and Science, Grant No.~20244028.

\newpage

\begin{appendix}

\section{SU(5) generators and structure constants \label{ap:su5g}}

The generators of SU(5) transformations, $\lambda_a$ ($a=1,\cdots 24$)
are 
\begin{eqnarray*}
&&\!\!\!\!\!\!\!\!
 \lambda_1\!\!=\!\!\left(\!
  \begin{array}{ccccc}
   {}^\cdot_1\! & \! {}^1_\cdot\! & \! {}^\cdot_\cdot\! & \!
    {}^\cdot_\cdot\! & \! {}^\cdot_\cdot\! \\ 
   {}^\cdot_\cdot\! & \! {}^\cdot_\cdot\! & \! {}^\cdot_\cdot\! & \!
    {}^\cdot_\cdot\! & \! {}^\cdot_\cdot\! \\ 
   {}^\cdot \! & \! {}^\cdot \! & \! {}^\cdot \! & \!
    {}^\cdot \! & \! {}^\cdot \!\\ 
  \end{array}\!\right)\!\! ,~
 \lambda_2\!\!=\!\!\left(\!
  \begin{array}{ccccc}
   {}^\cdot_{i}\! & \! {}^{-i}_{~\cdot}\! & \! {}^\cdot_\cdot\! & \!
    {}^\cdot_\cdot\! & \! {}^\cdot_\cdot\! \\ 
   {}^\cdot_\cdot\! & \! {}^\cdot_\cdot\! & \! {}^\cdot_\cdot\! & \!
    {}^\cdot_\cdot\! & \! {}^\cdot_\cdot\! \\ 
   {}^\cdot \! & \! {}^\cdot \! & \! {}^\cdot \! & \!
    {}^\cdot \! & \! {}^\cdot \! \\ 
  \end{array}\!\right)\!\! ,~
 \lambda_3\!\!=\!\!\left(\!
  \begin{array}{ccccc}
   {}^1_\cdot\! & \! {}^{~\cdot}_{-1}\! & \! {}^\cdot_\cdot\! & \!
    {}^\cdot_\cdot\! & \! {}^\cdot_\cdot\! \\ 
   {}^\cdot_\cdot\! & \! {}^\cdot_\cdot\! & \! {}^\cdot_\cdot\! & \!
    {}^\cdot_\cdot\! & \! {}^\cdot_\cdot\! \\ 
   {}^\cdot \! & \! {}^\cdot \! & \! {}^\cdot \! & \!
    {}^\cdot \! & \! {}^\cdot \! \\ 
  \end{array}\!\right)\!\! ,~
 \lambda_4\!\!=\!\!\left(\!
  \begin{array}{ccccc}
   {}^\cdot_\cdot\! & \! {}^\cdot_\cdot\! & \! {}^1_\cdot\! & \!
    {}^\cdot_\cdot\! & \! {}^\cdot_\cdot\! \\ 
   {}^1_\cdot\! & \! {}^\cdot_\cdot\! & \! {}^\cdot_\cdot\! & \!
    {}^\cdot_\cdot\! & \! {}^\cdot_\cdot\! \\ 
   {}^\cdot \! & \! {}^\cdot \! & \! {}^\cdot \! & \!
    {}^\cdot \! & \! {}^\cdot \! \\ 
  \end{array}\!\right)\!\! ,~
\\ &&\!\!\!\!\!\!\!\!
 \lambda_5\!\!=\!\!\left(\!
  \begin{array}{ccccc}
   {}^\cdot_\cdot\! & \! {}^\cdot_\cdot\! & \! {}^{-i}_{~\cdot}\! & \!
    {}^\cdot_\cdot\! & \! {}^\cdot_\cdot\! \\ 
   {}^i_\cdot\! & \! {}^\cdot_\cdot\! & \! {}^\cdot_\cdot\! & \!
    {}^\cdot_\cdot\! & \! {}^\cdot_\cdot\! \\ 
   {}^\cdot \! & \! {}^\cdot \! & \! {}^\cdot \! & \!
    {}^\cdot \! & \! {}^\cdot \! \\ 
  \end{array}\!\right)\!\! ,~~
 \lambda_6\!\!=\!\!\left(\!
  \begin{array}{ccccc}
   {}^\cdot_\cdot\! & \! {}^\cdot_\cdot\! & \! {}^\cdot_1\! & \!
    {}^\cdot_\cdot\! & \! {}^\cdot_\cdot\!  \\ 
   {}^\cdot_\cdot\! & \! {}^1_\cdot\! & \! {}^\cdot_\cdot\! & \!
    {}^\cdot_\cdot\! & \! {}^\cdot_\cdot\! \\ 
   {}^\cdot \! & \! {}^\cdot \! & \! {}^\cdot \! & \!
    {}^\cdot \! & \! {}^\cdot \! \\ 
  \end{array}\!\right)\!\! ,~~
 \lambda_7\!\!=\!\!\left(\!
  \begin{array}{ccccc}
   {}^\cdot_\cdot\! & \! {}^\cdot_\cdot\! & \! {}^{~\cdot}_{-i}\! & \!
    {}^\cdot_\cdot\! & \! {}^\cdot_\cdot\! \\ 
   {}^\cdot_\cdot\! & \! {}^i_\cdot\! & \! {}^\cdot_\cdot\! & \!
    {}^\cdot_\cdot\! & \! {}^\cdot_\cdot\! \\ 
   {}^\cdot \! & \! {}^\cdot \! & \! {}^\cdot \! & \!
    {}^\cdot \! & \! {}^\cdot \! \\ 
  \end{array}\!\right)\!\! ,~~
\\ &&\!\!\!\!\!\!\!\!
 \lambda_8\!\!=\!\!\textrm{${1\over \sqrt{3}}$}\left(\!
  \begin{array}{ccccc}
   {}^1_\cdot\! & \! {}^\cdot_1\! & \! {}^\cdot_\cdot\! & \!
    {}^\cdot_\cdot\! & \! {}^\cdot_\cdot\! \\ 
   {}^\cdot_\cdot\! & \! {}^\cdot_\cdot\! & \! {}^{-2}_{~\cdot}\! & \!
    {}^\cdot_\cdot\! & \! {}^\cdot_\cdot\! \\ 
   {}^\cdot \! & \! {}^\cdot \! & \! {}^\cdot \! & \!
    {}^\cdot \! & \! {}^\cdot \! \\ 
  \end{array}\!\right)\!\! ,~~
 \lambda_9\!\!=\!\!\left(\!
  \begin{array}{ccccc}
   {}^\cdot_\cdot\! & \! {}^\cdot_\cdot\! & \! {}^\cdot_\cdot\! & \!
    {}^1_\cdot\! & \! {}^\cdot_\cdot\! \\ 
   {}^\cdot_1\! & \! {}^\cdot_\cdot\! & \! {}^\cdot_\cdot\! & \!
    {}^\cdot_\cdot\! & \! {}^\cdot_\cdot\! \\ 
   {}^\cdot \! & \! {}^\cdot \! & \! {}^\cdot \! & \!
    {}^\cdot \! & \! {}^\cdot \! \\ 
  \end{array}\!\right)\!\! ,~~
 \lambda_{10}\!\!=\!\!\left(\!
  \begin{array}{ccccc}
   {}^\cdot_\cdot\! & \! {}^\cdot_\cdot\! & \! {}^\cdot_\cdot\! & \!
    {}^{-i}_{~\cdot}\! & \! {}^\cdot_\cdot\! \\ 
   {}^\cdot_{i}\! & \! {}^\cdot_\cdot\! & \! {}^\cdot_\cdot\! & \!
    {}^\cdot_\cdot\! & \! {}^\cdot_\cdot\! \\ 
   {}^\cdot \! & \! {}^\cdot \! & \! {}^\cdot \! & \!
    {}^\cdot \! & \! {}^\cdot \! \\ 
  \end{array}\!\right)\!\! ,~~
\\ &&\!\!\!\!\!\!\!\!
 \lambda_{11}\!\!=\!\!\left(\!
  \begin{array}{ccccc}
   {}^\cdot_\cdot\! & \! {}^\cdot_\cdot\! & \! {}^\cdot_\cdot\! & \!
    {}^\cdot_1\! & \! {}^\cdot_\cdot\! \\ 
   {}^\cdot_\cdot\! & \! {}^\cdot_1\! & \! {}^\cdot_\cdot\! & \!
    {}^\cdot_\cdot\! & \! {}^\cdot_\cdot\! \\ 
   {}^\cdot\! & \! {}^\cdot\! & \! {}^\cdot\! & \!
    {}^\cdot\! & \! {}^\cdot\! \\ 
  \end{array}\!\right)\!\! ,~
 \lambda_{12}\!\!=\!\!\left(\!
  \begin{array}{ccccc}
   {}^\cdot_\cdot\! & \! {}^\cdot_\cdot\! & \! {}^\cdot_\cdot\! & \!
    {}^{~\cdot}_{-i}\! & \! {}^\cdot_\cdot\! \\ 
   {}^\cdot_\cdot\! & \! {}^\cdot_i\! & \! {}^\cdot_\cdot\! & \!
    {}^\cdot_\cdot\! & \! {}^\cdot_\cdot\! \\ 
   {}^\cdot\! & \! {}^\cdot\! & \! {}^\cdot\! & \!
    {}^\cdot\! & \! {}^\cdot\! \\ 
  \end{array}\!\right)\!\! ,~
 \lambda_{13}\!\!=\!\!\left(\!
  \begin{array}{ccccc}
   {}^\cdot_\cdot\! & \! {}^\cdot_\cdot\!&\! {}^\cdot_\cdot\!&\!
    {}^\cdot_\cdot\!&\! {}^\cdot_\cdot\!\\ 
   {}^\cdot_\cdot\!&\! {}^\cdot_\cdot\!&\! {}^\cdot_1\!&\!
    {}^1_\cdot\!&\! {}^\cdot_\cdot\!\\ 
   {}^\cdot\!&\! {}^\cdot\!&\! {}^\cdot\!&\!
    {}^\cdot\!&\! {}^\cdot\!\\ 
  \end{array}\!\right)\!\! ,~
 \lambda_{14}\!\!=\!\!\left(\!
  \begin{array}{ccccc}
   {}^\cdot_\cdot\!&\! {}^\cdot_\cdot\!&\! {}^\cdot_\cdot\!&\!
    {}^\cdot_\cdot\!&\! {}^\cdot_\cdot\!\\ 
   {}^\cdot_\cdot\!&\! {}^\cdot_\cdot\!&\! {}^\cdot_i \!&\!
    {}^{-i}_{~\cdot}\!&\! {}^\cdot_\cdot\!\\ 
   {}^\cdot\!&\! {}^\cdot\!&\! {}^\cdot\!&\!
    {}^\cdot\!&\! {}^\cdot\\ 
  \end{array}\!\right)\!\! ,~
\\ &&\!\!\!\!\!\!\!\!
 \lambda_{15}\!\!=\!\!\textrm{${1\over \sqrt{6}}$}\left(\!
  \begin{array}{ccccc}
   {}^1_\cdot\!&\! {}^\cdot_1\!&\! {}^\cdot_\cdot\!&\!
    {}^\cdot_\cdot\!&\! {}^\cdot_\cdot\!\\ 
   {}^\cdot_\cdot\!&\! {}^\cdot_\cdot\!&\! {}^1_\cdot\!&\!
    {}^{~\cdot}_{-3}\!&\! {}^\cdot_\cdot\!\\ 
   {}^\cdot\!&\! {}^\cdot\!&\! {}^\cdot\!&\!
    {}^\cdot\!&\! {}^\cdot\\ 
  \end{array}\!\right)\!\! ,~~
 \lambda_{16}\!\!=\!\!\left(\!
  \begin{array}{ccccc}
   {}^\cdot_\cdot\!&\! {}^\cdot_\cdot\!&\! {}^\cdot_\cdot\!&\!
    {}^\cdot_\cdot\!&\! {}^1_\cdot\!\\ 
   {}^\cdot_\cdot\!&\! {}^\cdot_\cdot\!&\! {}^\cdot_\cdot\!&\!
    {}^\cdot_\cdot\!&\! {}^\cdot_\cdot\!\\ 
   {}^1\!&\! {}^\cdot\!&\! {}^\cdot\!&\!
    {}^\cdot\!&\! {}^\cdot\\ 
  \end{array}\!\right)\!\! ,~~
 \lambda_{17}\!\!=\!\!\left(\!
  \begin{array}{ccccc}
   {}^\cdot_\cdot\!&\! {}^\cdot_\cdot\!&\! {}^\cdot_\cdot\!&\!
    {}^\cdot_\cdot\!&\! {}^{-i}_{~\cdot}\!\\ 
   {}^\cdot_\cdot\!&\! {}^\cdot_\cdot\!&\! {}^\cdot_\cdot\!&\!
    {}^\cdot_\cdot\!&\! {}^\cdot_\cdot\!\\ 
   {}^i\!&\! {}^\cdot\!&\! {}^\cdot\!&\!
    {}^\cdot\!&\! {}^\cdot\\ 
  \end{array}\!\right)\!\! ,~~
\\ &&\!\!\!\!\!\!\!\!
 \lambda_{18}\!\!=\!\!\left(\!
  \begin{array}{ccccc}
   {}^\cdot_\cdot\!&\! {}^\cdot_\cdot\!&\! {}^\cdot_\cdot\!&\!
    {}^\cdot_\cdot\!&\! {}^\cdot_1\!\\ 
   {}^\cdot_\cdot\!&\! {}^\cdot_\cdot\!&\! {}^\cdot_\cdot\!&\!
    {}^\cdot_\cdot\!&\! {}^\cdot_\cdot\!\\ 
   {}^\cdot\!&\! {}^1\!&\! {}^\cdot\!&\!
    {}^\cdot\!&\! {}^\cdot\\ 
  \end{array}\!\right)\!\! ,~
 \lambda_{19}\!\!=\!\!\left(\!
  \begin{array}{ccccc}
   {}^\cdot_\cdot\!&\! {}^\cdot_\cdot\!&\! {}^\cdot_\cdot\!&\!
    {}^\cdot_\cdot\!&\! {}^{~\cdot}_{-i}\!\\ 
   {}^\cdot_\cdot\!&\! {}^\cdot_\cdot\!&\! {}^\cdot_\cdot\!&\!
    {}^\cdot_\cdot\!&\! {}^\cdot_\cdot\!\\ 
   {}^\cdot\!&\! {}^i\!&\! {}^\cdot\!&\!
    {}^\cdot\!&\! {}^\cdot\\ 
  \end{array}\!\right)\!\! ,~
 \lambda_{20}\!\!=\!\!\left(\!
  \begin{array}{ccccc}
   {}^\cdot_\cdot\!&\! {}^\cdot_\cdot\!&\! {}^\cdot_\cdot\!&\!
    {}^\cdot_\cdot\!&\! {}^\cdot_\cdot\!\\ 
   {}^\cdot_\cdot\!&\! {}^\cdot_\cdot\!&\! {}^\cdot_\cdot\!&\!
    {}^\cdot_\cdot\!&\! {}^1_\cdot\!\\ 
   {}^\cdot\!&\! {}^\cdot\!&\! {}^1\!&\!
    {}^\cdot\!&\! {}^\cdot\\ 
  \end{array}\!\right)\!\! ,~
 \lambda_{21}\!\!=\!\!\left(\!
  \begin{array}{ccccc}
   {}^\cdot_\cdot\!&\! {}^\cdot_\cdot\!&\! {}^\cdot_\cdot\!&\!
    {}^\cdot_\cdot\!&\! {}^\cdot_\cdot\!\\ 
   {}^\cdot_\cdot\!&\! {}^\cdot_\cdot\!&\! {}^\cdot_\cdot\!&\!
    {}^\cdot_\cdot\!&\! {}^{-i}_{~\cdot}\!\\ 
   {}^\cdot\!&\! {}^\cdot\!&\! {}^i\!&\!
    {}^\cdot\!&\! {}^\cdot\\ 
  \end{array}\!\right)\!\! ,~
\\ &&\!\!\!\!\!\!\!\!
 \lambda_{22}\!\!=\!\!\left(\!
  \begin{array}{ccccc}
   {}^\cdot_\cdot\!&\! {}^\cdot_\cdot\!&\! {}^\cdot_\cdot\!&\!
    {}^\cdot_\cdot\!&\! {}^\cdot_\cdot\!\\ 
   {}^\cdot_\cdot\!&\! {}^\cdot_\cdot\!&\! {}^\cdot_\cdot\!&\!
    {}^\cdot_\cdot\!&\! {}^\cdot_1\!\\ 
   {}^\cdot\!&\! {}^\cdot\!&\! {}^\cdot\!&\!
    {}^1\!&\! {}^\cdot\\ 
  \end{array}\!\right)\!\! ,~~
 \lambda_{23}\!\!=\!\!\left(\!
  \begin{array}{ccccc}
   {}^\cdot_\cdot\!&\! {}^\cdot_\cdot\!&\! {}^\cdot_\cdot\!&\!
    {}^\cdot_\cdot\!&\! {}^\cdot_\cdot\!\\ 
   {}^\cdot_\cdot\!&\! {}^\cdot_\cdot\!&\! {}^\cdot_\cdot\!&\!
    {}^\cdot_\cdot\!&\! {}^{~\cdot}_{-i}\!\\ 
   {}^\cdot\!&\! {}^\cdot\!&\! {}^\cdot\!&\!
    {}^i\!&\! {}^\cdot\\ 
  \end{array}\!\right)\!\! ,~~
 \lambda_{24}\!\!=\!\!\textrm{${1\over \sqrt{10}}$}\left(\!
  \begin{array}{ccccc}
   {}^1_\cdot\!&\! {}^\cdot_1\!&\! {}^\cdot_\cdot\!&\!
    {}^\cdot_\cdot\!&\! {}^\cdot_\cdot\!\\ 
   {}^\cdot_\cdot\!&\! {}^\cdot_\cdot\!&\! {}^1_\cdot\!&\!
    {}^\cdot_1\!&\! {}^\cdot_\cdot\!\\ 
   {}^\cdot\!&\! {}^\cdot\!&\! {}^\cdot\!&\!
    {}^\cdot\!&\! {}^{-4}\\ 
  \end{array}\!\right)\!\! ,
\end{eqnarray*}
where dots indicate 0.
Here $T_a ={1\over 2} \lambda_a$.
The $\lambda_a$ obey the following commutation relationship:
\begin{eqnarray}
 \left[\lambda_a, \lambda_b\right]
  \equiv \lambda_a \lambda_b -\lambda_b \lambda_a
   =2i f_{abc}\lambda_c .
 \label{commu}
\end{eqnarray}
The $f_{abc}$ are odd under the permutation of any pair of indices.
The nonzero values are tabulated in Table~\ref{fabc}.
\begin{table}[h]
\caption{The $f_{abc}$. There are 66 independent nonzero structure constants. 
\label{fabc}}
\vspace{-7mm}
\begin{center}
\begin{minipage}[t]{2.8cm}
\begin{eqnarray*}
 \begin{array}{lrl}
 a~b~c && f_{abc} \\ \hline 
 {}_1 ~{}_2 ~{}_3 && {}_1 \\
 {}_1 ~{}_4 ~{}_7 && {}_{1/2}\\
 {}_1 ~{}_5 ~{}_6 &{}_-\!\!\!\!\!& {}_{1/2}\\
 {}_1 ~{}_9 ~{}_{12} && {}_{1/2}\\
 {}_1 ~{}_{10}~{}_{11} &{}_-\!\!\!\!\!& {}_{1/2}\\
 {}_1~{}_{16}~{}_{19} &&{}_{1/2} \\
 {}_1~{}_{17}~{}_{18} &{}_-\!\!\!\!\!&{}_{1/2} \\
 {}_2~{}_4~{}_6&&{}_{1/2}\\
 {}_2~{}_5~{}_7&&{}_{1/2}\\
 {}_2~{}_9~{}_{11}&&{}_{1/2}\\
 {}_2~{}_{10}~{}_{12}&&{}_{1/2}\\
 {}_2~{}_{16}~{}_{18}&&{}_{1/2}\\
 {}_2~{}_{17}~{}_{19}&&{}_{1/2}\\
 {}_3~{}_4~{}_5&&{}_{1/2}\\
 {}_3~{}_6~{}_7&{}_-\!\!\!\!\!&{}_{1/2}\\
 {}_3~{}_9~{}_{10}&&{}_{1/2}\\
 {}_3~{}_{11}~{}_{12}&{}_-\!\!\!\!\!&{}_{1/2}\\
 \end{array}
\end{eqnarray*}
\end{minipage}
\begin{minipage}[t]{2.9cm}
\begin{eqnarray*}
 \begin{array}{|lrl}
 a~b~c && f_{abc} \\ \hline 
 {}_3~{}_{16}~{}_{17}&&{}_{1/2}\\
 {}_3~{}_{18}~{}_{19}&{}_-\!\!\!\!\!&{}_{1/2}\\
 {}_4 ~{}_5 ~{}_8 && {}_{\sqrt{3}/2} \\
 {}_4 ~{}_9 ~{}_{14} && {}_{1/2}\\
 {}_4 ~{}_{10} ~{}_{13} &{}_-\!\!\!\!\!& {}_{1/2}\\
 {}_4 ~{}_{16} ~{}_{21} && {}_{1/2}\\
 {}_4 ~{}_{17}~{}_{20} &{}_-\!\!\!\!\!& {}_{1/2}\\
 {}_5~{}_{9}~{}_{13} &&{}_{1/2} \\
 {}_5~{}_{10}~{}_{14}&&{}_{1/2} \\
 {}_5~{}_{16}~{}_{20}&&{}_{1/2}\\
 {}_5~{}_{17}~{}_{21}&&{}_{1/2}\\
 {}_6~{}_{7}~{}_{8}&&{}_{\sqrt{3}/2}\\
 {}_6~{}_{11}~{}_{14}&&{}_{1/2}\\
 {}_6~{}_{12}~{}_{13}&{}_-\!\!\!\!\!&{}_{1/2}\\
 {}_6~{}_{18}~{}_{21}&&{}_{1/2}\\
 {}_6~{}_{19}~{}_{20}&{}_-\!\!\!\!\!&{}_{1/2}\\
 {}_7~{}_{11}~{}_{13}&&{}_{1/2}\\
 \end{array}
\end{eqnarray*}
\end{minipage}
\begin{minipage}[t]{3cm}
\begin{eqnarray*}
 \begin{array}{|lrl}
 a~b~c && f_{abc} \\ \hline 
 {}_7~{}_{12}~{}_{14}&&{}_{1/2}\\
 {}_7~{}_{18}~{}_{20}&&{}_{1/2}\\
 {}_7~{}_{19}~{}_{21}&&{}_{1/2}\\
 {}_8 ~{}_9 ~{}_{10} && {}_{\sqrt{3}/6}\\
 {}_8 ~{}_{11} ~{}_{12} && {}_{\sqrt{3}/6}\\
 {}_8 ~{}_{13} ~{}_{14} &{}_-\!\!\!\!\!& {}_{\sqrt{3}/3}\\
 {}_8 ~{}_{16}~{}_{17} && {}_{\sqrt{3}/6}\\
 {}_8~{}_{18}~{}_{19} &&{}_{\sqrt{3}/6} \\
 {}_8~{}_{20}~{}_{21} &{}_-\!\!\!\!\!&{}_{\sqrt{3}/3} \\
 {}_9~{}_{10}~{}_{15}&&{}_{\sqrt{6}/3}\\
 {}_9~{}_{16}~{}_{23}&&{}_{1/2}\\
 {}_9~{}_{17}~{}_{22}&{}_-\!\!\!\!\!&{}_{1/2}\\
 {}_{10}~{}_{16}~{}_{22}&&{}_{1/2}\\
 {}_{10}~{}_{17}~{}_{23}&&{}_{1/2}\\
 {}_{11}~{}_{12}~{}_{15}&&{}_{\sqrt{6}/3}\\
 {}_{11}~{}_{18}~{}_{23}&&{}_{1/2}\\
 \end{array}
\end{eqnarray*}
\end{minipage}
\begin{minipage}[t]{3.1cm}
\begin{eqnarray*}
 \begin{array}{|lrl}
 a~b~c && f_{abc} \\ \hline 
 {}_{11}~{}_{19}~{}_{22}&{}_-\!\!\!\!\!&{}_{1/2}\\
 {}_{12}~{}_{18}~{}_{22}&&{}_{1/2}\\
 {}_{12} ~{}_{19} ~{}_{23} && {}_{1/2} \\
 {}_{13} ~{}_{14} ~{}_{15} && {}_{\sqrt{6}/3}\\
 {}_{13} ~{}_{20}~{}_{23} && {}_{1/2}\\
 {}_{13}~{}_{21}~{}_{22} &{}_-\!\!\!\!\!&{}_{1/2} \\
 {}_{14}~{}_{20}~{}_{22}&&{}_{1/2} \\
 {}_{14}~{}_{21}~{}_{23}&&{}_{1/2}\\
 {}_{15}~{}_{16}~{}_{17}&&{}_{\sqrt{6}/12}\\
 {}_{15}~{}_{18}~{}_{19}&&{}_{\sqrt{6}/12}\\
 {}_{15}~{}_{20}~{}_{21}&&{}_{\sqrt{6}/12}\\
 {}_{15}~{}_{22}~{}_{23}&{}_-\!\!\!\!\!&{}_{\sqrt{6}/4}\\
 {}_{16}~{}_{17}~{}_{24}&&{}_{\sqrt{10}/4}\\ 
 {}_{18} ~{}_{19} ~{}_{24} && {}_{\sqrt{10}/4} \\
 {}_{20}~{}_{21}~{}_{24} &&{}_{\sqrt{10}/4} \\
 {}_{22}~{}_{23}~{}_{24}&&{}_{\sqrt{10}/4}\\
 \end{array}
\end{eqnarray*}
\end{minipage}
\end{center}
\end{table}

The generators~(\ref{so5l}) are given in a matrix form for $\lambda_a$ as
\bea
  && \lambda_{\overline{1}}={\lambda_1 -\lambda_{22}\over \sqrt{2}}
     =\textrm{${1\over \sqrt{2}}$}\left(\begin{array}{ccccc}
        {}_1^\cdot \!&\! {}_\cdot^1 \!&\! {}_\cdot^\cdot \!&\! 
        {}_\cdot^\cdot \!&\! {}_\cdot^\cdot \\
        {}_\cdot^\cdot \!&\! {}_\cdot^\cdot \!&\! {}_\cdot^\cdot \!&\!
        {}_\cdot^\cdot \!&\! {}_{-1}^{~\cdot} \\
        {}^\cdot \!&\! {}^\cdot \!&\! {}^\cdot \!&\! {}^{-1} \!&\! {}^\cdot \\
        \end{array}\right) ,
\quad 
         \lambda_{\overline{2}}={\lambda_2 +\lambda_{23}\over \sqrt{2}}
     =\textrm{${1\over \sqrt{2}}$}\left(\begin{array}{ccccc}
        {}_i^\cdot \!&\! {}_{~\cdot}^{-i} \!&\! {}_\cdot^\cdot \!&\! 
        {}_\cdot^\cdot \!&\! {}_\cdot^\cdot \\
        {}_\cdot^\cdot \!&\! {}_\cdot^\cdot \!&\! {}_\cdot^\cdot \!&\!
        {}_\cdot^\cdot \!&\! {}_{-i}^{~\cdot} \\
        {}^\cdot \!&\! {}^\cdot \!&\! {}^\cdot \!&\! {}^{i} \!&\! {}^\cdot \\
        \end{array}\right) ,
\nonumber
\\
   && \lambda_{\overline{3}}={\lambda_4 -\lambda_{13}\over \sqrt{2}}
     =\textrm{${1\over \sqrt{2}}$}\left(\begin{array}{ccccc}
        {}_\cdot^\cdot \!&\! {}_\cdot^\cdot \!&\! {}_\cdot^1 \!&\! 
        {}_\cdot^\cdot \!&\! {}_\cdot^\cdot \\
        {}_\cdot^1 \!&\! {}_\cdot^\cdot \!&\! {}_{-1}^{~\cdot} \!&\!
        {}_{~\cdot}^{-1} \!&\! {}_\cdot^\cdot \\
        {}^\cdot \!&\! {}^\cdot \!&\! {}^\cdot \!&\! {}^\cdot \!&\! {}^\cdot\\
        \end{array}\right) ,
\quad 
     \lambda_{\overline{4}}={\lambda_5 -\lambda_{14}\over \sqrt{2}}
     =\textrm{${1\over \sqrt{2}}$}\left(\begin{array}{ccccc}
        {}_\cdot^\cdot \!&\! {}_\cdot^\cdot \!&\! {}_{~\cdot}^{-i} \!&\! 
        {}_\cdot^\cdot \!&\! {}_\cdot^\cdot \\
        {}_\cdot^i \!&\! {}_\cdot^\cdot \!&\! {}_{-i}^{~\cdot} \!&\!
        {}_{\cdot}^{i} \!&\! {}_\cdot^\cdot \\
        {}^\cdot \!&\! {}^\cdot \!&\! {}^\cdot \!&\! {}^\cdot \!&\! {}^\cdot\\
        \end{array}\right) ,   
\nonumber
\\
   && \lambda_{\overline{5}}={\lambda_6 -\lambda_{20}\over \sqrt{2}}
     =\textrm{${1\over \sqrt{2}}$}\left(\begin{array}{ccccc}
        {}_\cdot^\cdot \!&\! {}_\cdot^\cdot \!&\! {}_1^\cdot \!&\! 
        {}_\cdot^\cdot \!&\! {}_\cdot^\cdot \\
        {}_\cdot^\cdot \!&\! {}_\cdot^1 \!&\! {}_\cdot^{\cdot} \!&\!
        {}_{\cdot}^{\cdot} \!&\! {}_{~\cdot}^{-1} \\
        {}^\cdot \!&\! {}^\cdot \!&\! {}^{-1} \!&\! {}^\cdot \!&\! {}^\cdot\\
        \end{array}\right) ,
 \quad
       \lambda_{\overline{6}}={\lambda_7 -\lambda_{21}\over \sqrt{2}}
     =\textrm{${1\over \sqrt{2}}$}\left(\begin{array}{ccccc}
        {}_\cdot^\cdot \!&\! {}_\cdot^\cdot \!&\! {}_{-i}^{~\cdot} \!&\! 
        {}_\cdot^\cdot \!&\! {}_\cdot^\cdot \\
        {}_\cdot^\cdot \!&\! {}_\cdot^i \!&\! {}_\cdot^{\cdot} \!&\!
        {}_{\cdot}^{\cdot} \!&\! {}_{\cdot}^{i} \\
        {}^\cdot \!&\! {}^\cdot \!&\! {}^{-i} \!&\! {}^\cdot \!&\! {}^\cdot\\
        \end{array}\right) ,
\nonumber
\\
   && \lambda_{\overline{7}}={\lambda_{11} -\lambda_{16}\over \sqrt{2}}
     =\textrm{${1\over \sqrt{2}}$}\left(\begin{array}{ccccc}
        {}_\cdot^\cdot \!&\! {}_\cdot^\cdot \!&\! {}_\cdot^\cdot \!&\! 
        {}_1^\cdot \!&\! {}_{~\cdot}^{-1} \\
        {}_\cdot^\cdot \!&\! {}_1^\cdot \!&\! {}_\cdot^{\cdot} \!&\!
        {}_{\cdot}^{\cdot} \!&\! {}_{\cdot}^{\cdot} \\
        {}^{-1} \!&\! {}^\cdot \!&\! {}^{\cdot} \!&\! {}^\cdot \!&\! {}^\cdot\\
        \end{array}\right) ,
  \quad
        \lambda_{\overline{8}}={\lambda_{12} -\lambda_{17}\over \sqrt{2}}
     =\textrm{${1\over \sqrt{2}}$}\left(\begin{array}{ccccc}
        {}_\cdot^\cdot \!&\! {}_\cdot^\cdot \!&\! {}_\cdot^\cdot \!&\! 
        {}_{-i}^{~\cdot} \!&\! {}_{\cdot}^{i} \\
        {}_\cdot^\cdot \!&\! {}_i^\cdot \!&\! {}_\cdot^{\cdot} \!&\!
        {}_{\cdot}^{\cdot} \!&\! {}_{\cdot}^{\cdot} \\
        {}^{-i} \!&\! {}^\cdot \!&\! {}^{\cdot} \!&\! {}^\cdot \!&\! {}^\cdot\\
        \end{array}\right) ,
\nonumber
\\
  && 
  \lambda_{\overline{9}}
    = \textrm{${1\over \sqrt{2}}$}(\lambda_3 +\textrm{${\sqrt{6}\over 4}$}\lambda_{15}
      -\textrm{${\sqrt{10}\over 4}$}\lambda_{24}) 
    =\textrm{${1\over \sqrt{2}}$}\left(\begin{array}{ccccc}
        {}_\cdot^1 \!&\! {}_{-1}^{~\cdot} \!&\! {}_\cdot^\cdot \!&\! 
        {}_{\cdot}^{\cdot} \!&\! {}_{\cdot}^{\cdot} \\
        {}_\cdot^\cdot \!&\! {}_\cdot^\cdot \!&\! {}_\cdot^{\cdot} \!&\!
        {}_{-1}^{~\cdot} \!&\! {}_{\cdot}^{\cdot} \\
        {}^{\cdot} \!&\! {}^\cdot \!&\! {}^{\cdot} \!&\! {}^\cdot \!&\! {}^1\\
        \end{array}\right) ,
\nonumber
\\
  && 
  \lambda_{\overline{10}}
    = \textrm{${1\over \sqrt{2}}$}(\textrm{${\sqrt{3}\over 3}$}
   \lambda_8 +\textrm{${5\sqrt{6}\over 12}$}\lambda_{15}
      +\textrm{${\sqrt{10}\over 4}$}\lambda_{24}) 
    =\textrm{${1\over \sqrt{2}}$}\left(\begin{array}{ccccc}
        {}_\cdot^1 \!&\! {}_{1}^{\cdot} \!&\! {}_\cdot^\cdot \!&\! 
        {}_{\cdot}^{\cdot} \!&\! {}_{\cdot}^{\cdot} \\
        {}_\cdot^\cdot \!&\! {}_\cdot^\cdot \!&\! {}_\cdot^{\cdot} \!&\!
        {}_{-1}^{~\cdot} \!&\! {}_{\cdot}^{\cdot} \\
        {}^{\cdot} \!&\! {}^\cdot \!&\! {}^{\cdot} \!&\! {}^\cdot \!&\! {}^{-1}\\
        \end{array}\right) .
        \label{matrixform}
\eea
The other generators of SU(5) are written as
\bea
  && \lambda_{\overline{11}}={\lambda_1 +\lambda_{22}\over \sqrt{2}}
     =\textrm{${1\over \sqrt{2}}$}\left(\begin{array}{ccccc}
        {}_1^\cdot \!&\! {}_\cdot^1 \!&\! {}_\cdot^\cdot \!&\! 
        {}_\cdot^\cdot \!&\! {}_\cdot^\cdot \\
        {}_\cdot^\cdot \!&\! {}_\cdot^\cdot \!&\! {}_\cdot^\cdot \!&\!
        {}_\cdot^\cdot \!&\! {}_{1}^{~\cdot} \\
        {}^\cdot \!&\! {}^\cdot \!&\! {}^\cdot \!&\! {}^{1} \!&\! {}^\cdot \\
        \end{array}\right) ,
\quad 
         \lambda_{\overline{12}}={\lambda_2 -\lambda_{23}\over \sqrt{2}}
     =\textrm{${1\over \sqrt{2}}$}\left(\begin{array}{ccccc}
        {}_i^\cdot \!&\! {}_{~\cdot}^{-i} \!&\! {}_\cdot^\cdot \!&\! 
        {}_\cdot^\cdot \!&\! {}_\cdot^\cdot \\
        {}_\cdot^\cdot \!&\! {}_\cdot^\cdot \!&\! {}_\cdot^\cdot \!&\!
        {}_\cdot^\cdot \!&\! {}_{i}^{~\cdot} \\
        {}^\cdot \!&\! {}^\cdot \!&\! {}^\cdot \!&\! {}^{-i} \!&\! {}^\cdot \\
        \end{array}\right) ,
\nonumber
\\
   && \lambda_{\overline{13}}={\lambda_4 +\lambda_{13}\over \sqrt{2}}
     =\textrm{${1\over \sqrt{2}}$}\left(\begin{array}{ccccc}
        {}_\cdot^\cdot \!&\! {}_\cdot^\cdot \!&\! {}_\cdot^1 \!&\! 
        {}_\cdot^\cdot \!&\! {}_\cdot^\cdot \\
        {}_\cdot^1 \!&\! {}_\cdot^\cdot \!&\! {}_{1}^{~\cdot} \!&\!
        {}_{~\cdot}^{1} \!&\! {}_\cdot^\cdot \\
        {}^\cdot \!&\! {}^\cdot \!&\! {}^\cdot \!&\! {}^\cdot \!&\! {}^\cdot\\
        \end{array}\right) ,
\quad 
     \lambda_{\overline{14}}={\lambda_5 +\lambda_{14}\over \sqrt{2}}
     =\textrm{${1\over \sqrt{2}}$}\left(\begin{array}{ccccc}
        {}_\cdot^\cdot \!&\! {}_\cdot^\cdot \!&\! {}_{~\cdot}^{-i} \!&\! 
        {}_\cdot^\cdot \!&\! {}_\cdot^\cdot 
      \\
        {}_\cdot^{i} \!&\! {}_\cdot^\cdot \!&\! {}_{i}^{~\cdot} \!&\!
        {}_{\cdot}^{-i} \!&\! {}_\cdot^\cdot 
      \\
        {}^\cdot \!&\! {}^\cdot \!&\! {}^\cdot \!&\! {}^\cdot \!&\! {}^\cdot
      \\
        \end{array}\right) ,   
\nonumber
\\
   && \lambda_{\overline{15}}={\lambda_6 +\lambda_{20}\over \sqrt{2}}
     =\textrm{${1\over \sqrt{2}}$}\left(\begin{array}{ccccc}
        {}_\cdot^\cdot \!&\! {}_\cdot^\cdot \!&\! {}_1^\cdot \!&\! 
        {}_\cdot^\cdot \!&\! {}_\cdot^\cdot \\
        {}_\cdot^\cdot \!&\! {}_\cdot^1 \!&\! {}_\cdot^{\cdot} \!&\!
        {}_{\cdot}^{\cdot} \!&\! {}_{~\cdot}^{1} \\
        {}^\cdot \!&\! {}^\cdot \!&\! {}^{1} \!&\! {}^\cdot \!&\! {}^\cdot\\
        \end{array}\right) ,
 \quad
       \lambda_{\overline{16}}={\lambda_7 +\lambda_{21}\over \sqrt{2}}
     =\textrm{${1\over \sqrt{2}}$}\left(\begin{array}{ccccc}
        {}_\cdot^\cdot \!&\! {}_\cdot^\cdot \!&\! {}_{-i}^{~\cdot} \!&\! 
        {}_\cdot^\cdot \!&\! {}_\cdot^\cdot 
   \\
        {}_\cdot^\cdot \!&\! {}_\cdot^i \!&\! {}_\cdot^{\cdot} \!&\!
        {}_{\cdot}^{\cdot} \!&\! {}_{\cdot}^{-i} 
   \\
        {}^\cdot \!&\! {}^\cdot \!&\! {}^{i} \!&\! {}^\cdot \!&\! {}^\cdot\\
        \end{array}\right) ,
\nonumber
\\
   && \lambda_{\overline{17}}={\lambda_{11} +\lambda_{16}\over \sqrt{2}}
     =\textrm{${1\over \sqrt{2}}$}\left(\begin{array}{ccccc}
        {}_\cdot^\cdot \!&\! {}_\cdot^\cdot \!&\! {}_\cdot^\cdot \!&\! 
        {}_1^\cdot \!&\! {}_{~\cdot}^{1} \\
        {}_\cdot^\cdot \!&\! {}_1^\cdot \!&\! {}_\cdot^{\cdot} \!&\!
        {}_{\cdot}^{\cdot} \!&\! {}_{\cdot}^{\cdot} \\
        {}^{1} \!&\! {}^\cdot \!&\! {}^{\cdot} \!&\! {}^\cdot \!&\! {}^\cdot\\
        \end{array}\right) ,
  \quad
        \lambda_{\overline{18}}={\lambda_{12} +\lambda_{17}\over \sqrt{2}}
     =\textrm{${1\over \sqrt{2}}$}\left(\begin{array}{ccccc}
        {}_\cdot^\cdot \!&\! {}_\cdot^\cdot \!&\! {}_\cdot^\cdot \!&\! 
        {}_{-i}^{~\cdot} \!&\! {}_{\cdot}^{-i} \\
        {}_\cdot^\cdot \!&\! {}_i^\cdot \!&\! {}_\cdot^{\cdot} \!&\!
        {}_{\cdot}^{\cdot} \!&\! {}_{\cdot}^{\cdot} \\
        {}^{i} \!&\! {}^\cdot \!&\! {}^{\cdot} \!&\! {}^\cdot \!&\! {}^\cdot\\
        \end{array}\right) ,
\nonumber
\\
  && 
  \lambda_{\overline{19}}
    = \textrm{${1\over \sqrt{2}}$}(\lambda_3 -\textrm{${\sqrt{6}\over 4}$}\lambda_{15}
      +\textrm{${\sqrt{10}\over 4}$}\lambda_{24}) 
    =\textrm{${1\over \sqrt{2}}$}\left(\begin{array}{ccccc}
        {}_\cdot^1 \!&\! {}_{-1}^{~\cdot} \!&\! {}_\cdot^\cdot \!&\! 
        {}_{\cdot}^{\cdot} \!&\! {}_{\cdot}^{\cdot} \\
        {}_\cdot^\cdot \!&\! {}_\cdot^\cdot \!&\! {}_\cdot^{\cdot} \!&\!
        {}_{1}^{~\cdot} \!&\! {}_{\cdot}^{\cdot} \\
        {}^{\cdot} \!&\! {}^\cdot \!&\! {}^{\cdot} \!&\! {}^\cdot \!&\! {}^{-1}\\
        \end{array}\right) ,
\nonumber
\\
  && 
  \lambda_{\overline{20}}
    = \textrm{${1\over \sqrt{10}}$}(\textrm{${5\sqrt{3}\over 3}$}
   \lambda_8 -\textrm{${5\sqrt{6}\over 12}$}\lambda_{15}
      -\textrm{${\sqrt{10}\over 4}$}\lambda_{24}) 
    =\textrm{${1\over \sqrt{10}}$}\left(\begin{array}{ccccc}
        {}_\cdot^1 \!&\! {}_{1}^{\cdot} \!&\! {}_\cdot^\cdot \!&\! 
        {}_{\cdot}^{\cdot} \!&\! {}_{\cdot}^{\cdot} 
    \\
        {}_\cdot^\cdot \!&\! {}_\cdot^\cdot \!&\! {}_\cdot^{-4} \!&\!
        {}_{1}^{~\cdot} \!&\! {}_{\cdot}^{\cdot} 
    \\
        {}^{\cdot} \!&\! {}^\cdot \!&\! {}^{\cdot} \!&\! {}^\cdot \!&\! {}^{1}\\
        \end{array}\right) ,
\nonumber
\\
  && \lambda_9 , ~ \lambda_{10}, ~\lambda_{18}, ~\lambda_{19} .
\eea

\end{appendix}

\vspace{4ex}



\end{document}